\documentclass[11pt,dvips]{article}
\usepackage{graphicx}
\usepackage{amsmath}
\usepackage{amssymb}
\usepackage{latexsym}
\usepackage{color}
\begin{document}
\thispagestyle{empty}
\begin{center}
\vspace{1.8cm}
%%%%%%%%%%%%%%%%%%%%%%%%%%%%%%%%%%%%%%%%%%%%%%%%%%%%%%%%%%%%%%%%%%%%%%%%%%%%%%%%%%%%%%%%%%%%%%%%%%%%%%%%%%%%%%%%%%%%%%%
 {\large \bf A recursive approach for geometric quantifiers of quantum correlations in multiqubit Schr\"odinger cat states}\\
 %%%%%%%%%%%%%%%%%%%%%%%%%%%%%%%%%%%%%%%%%%%%%%%%%%%%%%%%%%%%%%%%%%%%%%%%%%%%%%%%%%%%%%%%%%%%%%%%%%%%%%%%%%%%%%%%%%%%%

\vspace{1.5cm} {\bf M. Daoud}$^{a,b,c}${\footnote { email: {\sf
m$_{-}$daoud@hotmail.com}}}, {\bf R. Ahl Laamara}$^{d,e}$ {\footnote
{ email: {\sf ahllaamara@gmail.com}}}  and {\bf S. Seddik}$^{d}$
{\footnote { email: {\sf sanaa1.sanaa@hotmail.fr}}} \\
\vspace{0.5cm}
$^{a}${\it Max Planck Institute for the Physics of Complex Systems, Dresden, Germany}\\[0.5em]
$^{b}${\it Abdus Salam  International Centre for Theoretical Physics, Trieste, Italy}\\[0.5em]
$^{c}${\it Department of Physics, Faculty of Sciences,  University Ibnou Zohr, Agadir , Morocco}\\[0.5em]
$^{d}${\it LPHE-Modeling and Simulation, Faculty  of Sciences, Rabat, Morocco}\\[0.5em]
$^{e}${\it Centre of Physics and Mathematics,  Rabat, Morocco}\\[0.5em]

\vspace{3cm} {\bf Abstract}
\end{center}
\baselineskip=18pt
\medskip

A recursive approach  to determine the Hilbert-Schmidt measure of
pairwise quantum discord in a special class of symmetric states of
$k$ qubits is presented. We especially focus on the reduced states
of $k$ qubits obtained from a balanced  superposition of symmetric
$n$-qubit states (multiqubit Schr\"odinger cat states) by tracing
out $n-k$ particles  $(k=2,3, \cdots ,n-1)$. Two pairing  schemes
are considered. In the first one, the geometric discord measuring
the correlation between one qubit and the party grouping $(k-1)$
qubits is explicitly derived. This uses recursive  relations
between the Fano-Bloch correlation matrices associated with
subsystems comprising  $k$, $k-1$, $\cdots$ and  $2$ particles. A
detailed analysis is given for two, three and four qubit systems. In
the second scheme, the subsystem comprising the $(k-1)$ qubits is
mapped into a system of two logical qubits. We show that these two
bipartition  schemes are equivalents in  evaluating  the pairwise
correlation in multi-qubits systems. The explicit expressions of
classical states presenting zero discord are derived.

\newpage
%%%%%%%%%%%%%%%%%%%%%%%%%%%%%%%%%%%%%%%%%%
\section{Introduction}
%%%%%%%%%%%%%%%%%%%%%%%%%%%%%%%%%%%%%%%%%%%%

Quantum correlations in multipartite systems have generated  a lot of interest
during the last two decades \cite{Horodecki-RMP-2009,
Guhne,vedral}. This is essentially motivated by their promising
 applications in the field of  quantum information
 such as implementing quantum cryptographic protocols, speeding up quantum computing algorithms
and many more quantum tasks (see for instance \cite{NC-QIQC-2000,Vedral-RMP-2002}). An important issue in investigating quantum correlations
concerns  the appropriate  measure to decide
about quantumness in a given quantum system and to separate between
classical and quantum states. The characterization of quantum correlations  is necessary  in order to exploit
their advantages, in
an efficient way,   in
 the context of  quantum information processing such as   quantum teleportation \cite{Bennett1}, superdense coding \cite{Bennett2}
and quantum key distribution \cite{Ekert}. Several methods and different measures of quantum correlation were exhaustively discussed in the literature
from various perspectives  and for many purposes (for a recent review see \cite{vedral}).
They can be classified in two main categories: entropic based measures and geometric quantifiers or
norm based measures. Entanglement of formation, linear
entropy, relative entropy and quantum discord \cite{Rungta,Ben3,Wootters,Coffman,Vedral-et-al,Ollivier-PRL88-2001}
constitute  familiar entropic quantifiers of correlations. Probably, quantum discord,
which goes beyond entanglement,
is the most prominent of these correlations. It has been the
subject of intensive studies during the last decade.  It was originally defined as the difference
between two quantum analogues of the classical mutual information \cite{Vedral-et-al,Ollivier-PRL88-2001}.
The explicit evaluation of based entropy measures require optimization procedures
which are in general very complicated to achieve and constitute the main obstacle in order to get
computable expressions of quantum correlations. To overcome such difficulties,
geometric measures, especially ones based on Hilbert-Schmidt norm, were considered
to formulate a geometric variant of quantum discord  \cite{Dakic2010}. The Hilbert-Schmidt distance
 was used to quantify classical correlations \cite{Bellomo1,Bellomo2}. We notice  that the measure of quantum and classical correlations
 in bipartite systems   can be  also evaluated  through
the 1-norm distance (trace distance) \cite{Dajka,Paula1,Aaronson,Paula2}.

In other hand, the extension of Hilbert-Schmidt measure of quantum
discord to  $d$-dimensional quantum systems (qudits) was reported
in \cite{Rana,Hassan,Rau} (see also \cite{Zhou} and references
quoted therein). It must be emphasized that this higher dimensional
extension can be adapted to understand the pairwise quantum
correlations in multi-qubit systems.  Indeed, geometric quantum discord based on the
Hilbert-Schmidt norm turns out to be more tractable, in multi-qubit systems, from a
computational point of view than entropic based measures. In this
sense, we employ the approach by Dakic et al \cite{Dakic2010} to
investigate the quantum correlations in mixed multi-qubit states. Specifically, we shall
consider a balanced superposition of  symmetric multi-qubit states in which the symmetry properties offer drastic simplification
in evaluating quantum correlations.

This paper is organized as follows. In section 2, we discuss the
relevance of symmetric multi-qubit ($n$-qubits) states in defining
Schr\"odinger cat states. We shall essentially focus on balanced
superpositions, symmetric or antisymmetric under the parity
transformation, which coincide with even and odd spin atomic coherent
states. A special attention, in section 3, is devoted to reduced states
describing subsystems containing $k$ qubits ($k = 2, 3,
\cdots, n-1$) obtained by tracing out $n-k$ qubit from a $n$-qubit
Schr\"odinger cat state. This trace procedure gives rise to states
called extended $X$ states. The algebraic structure of such states
 provides a nice prescription to evaluate the quantum
correlation based on Hilbert-Schmidt (geometric quantum discord)
between one qubit and $(k-1)$ qubits contained  in a mixed $k$-qubit
state. This procedure is explicitly described in section 4. We consider in detail
the cases of two and three qubit systems.  We develop  the general
method to determine analytically geometric discord in mixed
$k$-qubits states. We also derive the explicit forms of classical
(zero discord) states.
In section 5, we introduce another scheme according to which the second part
of the system containing  $k-1$ qubits is mapped into two logical qubits. In this picture the
whole system reduces to a two qubit system. Remarkably, the geometric measure
of quantum discord obtained,  in this second scheme, coincides with one derived
in the first bi-partition scheme (section 4).  As illustration, a detailed analysis is given for $k=3$
and $k=4$. The method developed in this
paper which extends the geometric measure of two-qubit $X$ states to
embrace $k$-qubit $X$ states is useful in investigating the
global pairwise  correlation in multipartite qubit systems.
 Concluding
remarks close this paper.

%%%%%%%%%%%%%%%%%%%%%%%%%%%%%%%%%%%%%%%%%%%%%%%%%%%%%%%%%%%%%%%%%%%%%%%%
\section{Symmetric multi-qubit systems}
%%%%%%%%%%%%%%%%%%%%%%%%%%%%%%%%%%%%%%%%%%%%%%%%%%%%%%%%%%%%%%%%%%%%%%%%

The multi-qubit symmetric states were shown relevant  for different purposes  in quantum information
science~\cite{solano,
mixed,usa1,usa2,markham1,markham2,gebastin,markham3}. In this paper,  we
shall mainly focus on an ensemble of $n$ spin-$1/2$ prepared in even and
 odd spin coherent states.

%%%%%%%%%%%%%%%%%%%%%%%%%%%%%%%%%%%%%%%%%%%%%%%%%%%%%%%%%%%%%%%%%%%%%%%
\subsection{Spin coherent as  symmetric multi-qubit systems}
%%%%%%%%%%%%%%%%%%%%%%%%%%%%%%%%%%%%%%%%%%%%%%%%%%%%%%%%%%%%%%%%%%%%%%%%

We consider $n$ identical qubits. Each qubit lives in a
2-dimensional Hilbert space
$\mathcal{H}=\mathrm{span}\{\vert{0}\rangle,\vert{1}\rangle\}$. The
Hilbert space of the $n$-qubit system is given  by $n$ tensored
copies of $\mathcal{H}$
$$\mathcal{H}_n := \mathcal{H}^{\otimes n}.$$
Among the multi-partite states in $\mathcal{H}_n$, multi-qubit
states obeying exchange symmetry are of special interest from
experimental as well as mathematical point of views. An arbitrary
symmetric $n$-qubit state is commonly represented in either
Majorana~\cite{Maj32} or  Dicke ~\cite{Dic54}
representation. Any multi-qubit state, invariant under the exchange
symmetry, is specified in the Majorana description by the state (up
to a normalization factor)
\begin{equation}\label{psisym}
|\psi_s\rangle = \frac{1}{n!} \sum_{\sigma \in {\cal S}_n} \vert {\eta_{\sigma(1)},\ldots,\eta_{\sigma(n)}}\rangle,
\end{equation}
where each single qubit state is $|\eta_i\rangle \equiv (1 + \eta_i
\bar\eta_i)^{-\frac{1}{2}}(|0\rangle + \eta_i |1\rangle)~
(i=1,\ldots, n$; the bar stands for complex conjugation)
%$|\eta_i\rangle \equiv \alpha_i |0\rangle + \beta_i |1\rangle$ ($i=1,\ldots, n$)
 and
the sum is over the elements of the permutation group ${\cal S}_n$
of $n$ objects. In Equation (\ref{psisym}), the vector
$\vert{\eta_{\sigma(1)},\ldots,\eta_{\sigma(n)}}\rangle$ stands for
the tensor product
$\vert{\eta_{\sigma(1)}}\rangle\otimes\ldots\otimes\vert{\eta_{\sigma(n)}}\rangle$.
The totally symmetric $n$-qubit states can be also formulated  in
Dike representation. The symmetric Dicke states with $k$ excitations
are defined by ~\cite{Dic54}
%\begin{equation}
%\label{Dicke}
%\vert {D_n(k)} \rangle = \frac{1}{\sqrt{C_n^k}} \sum_{\sigma} \vert{\underbrace{0\ldots 0}_{n-k}\underbrace{1\ldots 1}_{k}}\rangle,
%\end{equation}
\begin{equation}
\label{Dicke}
\vert n,k \rangle = \sqrt{\frac{k!(n-k)!}{n!}} \sum_{\sigma \in {\cal S}_n} \vert{\underbrace{0,\ldots, 0}_{n-k},\underbrace{1,\ldots, 1}_{k}}\rangle,
\end{equation}
which generate an orthonormal basis of the  symmetric Hilbert subspace of dimension $(n+1)$. Therefore, permutation invariance,  in symmetric
multi-qubit states, implies a  restriction to $n+1$ dimensional subspace from
the entire  $2^n$ dimensional Hilbert space. The Dicke states (\ref{Dicke}) constitute a special subset of
the symmetric multi-qubit states (\ref{psisym}) corresponding to the
situation where the first $k$-qubit are such that $\eta_i = 0$ for
$i = 0,1, \ldots k$ and the remaining qubits are in the states
$\vert \eta_i = 1 \rangle $ with $i = k+1, \ldots , n$.  Any symmetric state  $|\psi_s \rangle$ (\ref{psisym}) can be
expanded in terms of Dicke states (\ref{Dicke}) as follows
\begin{equation}
\label{psisymD}
|\psi_s\rangle = \frac{1}{n!} \sum_{k=0}^n c_k ~|n,k \rangle,
\end{equation}
where  the $c_k$ ($k=0,\ldots,n$) stand for the complex expansion
coefficients. In particular, when the qubit are all identical
($\eta_i = \eta$ for all qubits), it is simply verified that the
coefficients $c_k$ are given by
\begin{equation}
\label{ck} c_k = n!  \sqrt{\frac{n!}{k!(n-k)!}} \frac{\eta^{k}}{(1 +
\eta\bar\eta)^{\frac{n}{2}}}
\end{equation}
and the symmetric multi-qubit states  (\ref{psisym}) write
\begin{equation}\label{psisym-cs}
|\psi_s\rangle :=  \vert n, \eta \rangle = (1 +
\eta\bar\eta)^{-\frac{n}{2}} \sum_{k=0}^n \sqrt{\frac{n!}{k!(n-k)!}}
\eta^{k}  ~|n,k \rangle,
\end{equation}
which are exactly the  $j=\frac{n}{2}$-spin coherent states (for
more details see for instance \cite{daoud2}). In particular, the
state $\vert n, \eta \rangle $ can be identified for $n=1$ with
spin-$\frac{1}{2}$ coherent state with $|0\rangle \equiv
|\frac{1}{2}, -\frac{1}{2}\rangle$ and $|1\rangle \equiv
|\frac{1}{2} ,+ \frac{1}{2}\rangle$)

%%%%%%%%%%%%%%%%%%%%%%%%%%%%%%%%%%%%%%%%%%%%%%%%%%%%%%%%%%%%%%%%%%%%%%
\subsection{Multi-qubit "Shro\"odinger cat" states}
%%%%%%%%%%%%%%%%%%%%%%%%%%%%%%%%%%%%%%%%%%%%%%%%%%%%%%%%%%%%%%%%%%%%%%

The prototypical multi-qubit "Schr\"odinger cat" states, we consider
in this work, are defined  as a balanced superpositions of the
$n$-qubit states $\vert n, \eta \rangle$ and $\vert n, -\eta
\rangle$ given by (\ref{psisym-cs}). They write
\begin{equation}\label{ncs}
 \vert \eta, n , m \rangle  =  {\cal N}( \vert n, \eta \rangle + e^{im\pi} \vert n, - \eta
 \rangle)
\end{equation}
where
$$ \vert n, \pm \eta \rangle = \vert \pm \eta \rangle \otimes \vert \pm \eta \rangle \cdots \otimes \vert \pm \eta \rangle, $$
and the integer $m \in \mathbb{Z}$ takes the values $m = 0 ~({\rm
mod}~2)$ and $m = 1~ ({\rm mod}~2)$. The normalization factor ${\cal
N}$ is
$$ {\cal N}= \big[ 2 + 2 p^{n} \cos m \pi\big]^{-1/2}$$
where  $p$ denotes the overlap between the states $\vert \eta
\rangle$ and $\vert  -\eta \rangle$. It is given by
\begin{equation}\label{overlap}
 p = \langle \eta \vert  - \eta \rangle = \frac{1 - \bar\eta \eta}{1 + \bar\eta \eta}.
\end{equation}
 Experimental creation of cat states
comprising multiple particles was reported in the literature
\cite{Wineland,Haroche}. Due to their experimental implementation,
"Schr\"odinger cat" states are expected to be an useful resource for
quantum computing as well as quantum communications. Also, in view
of their mathematical elegance, multi-qubit states obeying exchange
symmetry offer drastic simplification  in investigating various
aspects of quantum correlations in particular the geometric measure
of quantum discord as we shall discuss in the present work.
Furthermore,  the multi-qubit symmetric states (\ref{ncs}) include
Greenberger-Horne-Zeilinger(${\rm GHZ}$)~\cite{ghz}, ${\rm W}$
\cite{Dur00} and Dicke states~\cite{Dic54}. The multi-qubits states
$ \vert n, \eta , 0 \rangle$ ($m=0 ~ {\rm mod}~ 2$) and $ \vert n,
\eta , 1 \rangle$ ($m=1 ~ {\rm mod}~ 2$) behave like a multipartite
state of Greenberger-Horne-Zeilinger (${\rm GHZ}$) type \cite{ghz}
in the limiting case $p \rightarrow 0 $. Indeed,
 the  states $|\eta \rangle $ and $|
-\eta \rangle $ approach orthogonality and an orthogonal basis can
be defined such that $\vert {\bf 0}\rangle\equiv \vert \eta \rangle$
and $\vert{\bf 1}\rangle \equiv \vert  -\eta \rangle$. Thus, the
state $\vert n , \eta, m \rangle$ becomes of ${\rm GHZ}$-type:
\begin{equation}
\vert \eta , n, m \rangle \sim \vert {\rm GHZ}\rangle_{n} = \frac
1{\sqrt{2}}(\vert {\bf 0}\rangle \otimes |{\bf 0}\rangle \otimes
        \cdots \otimes\vert {\bf 0}\rangle
    +e^{i m \pi}\vert {\bf 1}\rangle \otimes
    \vert {\bf 1}\rangle \otimes \cdots \otimes
\vert {\bf 1}\rangle).\label{GHZ}
\end{equation}
Also, in the special situation where the overlap $p$ tends to unity
($p \rightarrow 1$ or $ \eta \rightarrow 0$ ), the state $\vert \eta
, n, m = 0 ~({\rm mod}~ 2) \rangle$ (\ref{ncs}) reduces to ground
state of a collection of $n$ qubits
\begin{equation}
\vert 0,  n , 0 ~({\rm mod}~ 2) \rangle \sim  \vert 0
\rangle \otimes\vert 0 \rangle \otimes \cdots \otimes \vert
0 \rangle,
\end{equation}
and  it is simple to check that the state $\vert \eta , 0 , 1 ~({\rm mod}~ 2) \rangle$  becomes
a multipartite state of $W$ type~\cite{Dur00}
\begin{equation}
\vert 0,  n ,  1 ~({\rm mod}~ 2) \rangle \sim \vert\text{\rm
W}\rangle_{n}
    = \frac{1}{\sqrt{n}}(\vert 1 \rangle \otimes\vert 0\rangle \otimes \cdots\otimes
       \vert 0 \rangle  +\vert 0 \rangle \otimes\vert 1 \rangle \otimes\ldots\otimes \vert 0 \rangle
      +\cdots
   + \vert  0 \rangle \otimes\vert 0 \rangle  \otimes \cdots\otimes \vert 1 \rangle)~.
\label{Wstate}
\end{equation}
It is clear that the Schr\"odinger cat  states  $\vert \eta, n , m =
0 ~({\rm mod}~2) \rangle$  include the  ${\rm GHZ}_{n}$ states $(p
\rightarrow 0)$. In other hand, the states
 $\vert \eta, n , m = 1 ~({\rm mod}~2) \rangle$, constitute an interpolation between
two special classes of multi-qubits states: $\vert\text{\rm
GHZ}\rangle_{n}$ type corresponding to  $p \rightarrow 0$ and states
of $\vert\text{\rm W}\rangle_{n}$ type obtained in the special case
where $p \rightarrow 1$.

%%%%%%%%%%%%%%%%%%%%%%%%%%%%%%%%%%%%%%%%%%%%%%%%%%%%%%%%%%%%%%%%%%
\section{Multi-partite quantum correlations.}
%%%%%%%%%%%%%%%%%%%%%%%%%%%%%%%%%%%%%%%%%%%%%%%%%%%%%%%%%%%%%%%%%%%%%

The structure of multipartite correlations within multi-qubit
quantum systems is a challenging and daunting  task. With the growth of number of qubits, there are numerous ways in splitting the entire
 system to characterize how the particles are correlated. Obviously,
the bipartite splitting of the whole system is not sufficient to
capture the essential of quantum correlation existing in a
multi-qubit system. However, it must be noticed that the pairwise
decomposition of total correlation offers a good alternative  to
evaluate the amount of all correlations existing in a multipartite
system. In this paper, we approach the problem of analyzing
$n$-qubit correlation using only bipartite measures. Toward this
end, we consider first the correlation between one qubit with the
remaining $(n-1)$ qubits in the state (\ref{ncs}). Thus, the pure
density matrix of the symmetric $n$-qubit system writes
$$ \rho_{n}\equiv \vert \eta, n , m \rangle\langle  \eta, n , m  \vert := \rho_{1\vert 23\ldots n} $$
Furthermore, after removing $k=1, 2, \cdots, n-2$ particles from the
$n$-qubit system, the reduced density matrix $\rho_{n-k}$ can be
bi-partitioned in two subsystem, one comprises one qubit and the
remaining $(n-k-1)$ qubits are contained in the second subsystem. In this
manner, a bipartite measure characterize the pairwise correlation
between the two subsystem. This offers a reasonable scheme to
characterize  the total amount of quantum correlation defined as the sum
of the quantum correlations for all possible bi-partitions \cite{Fanchini2}.
%by the
%quantity
%\begin{equation} Q = \sum_{k=2}^{n} Q
%(\rho_{1\vert 23\ldots k})
%\end{equation}
%where $Q$ stands for a measure of quantum correlation in a bipartite
%system. It is interesting in the particular case of a tripartite
%state, one recovers the definition of multipartite quantum
%correlations introduced in \cite{Fanchini2}.

In this paper, we shall employ this picture to estimate the
geometric measure of quantum discord ($D_g$) in the
symmetric multi-qubit system of the form (\ref{ncs}). We give a
detailed analysis for two qubit and three qubit subsystems.  From these two specific cases,  we give a general
algorithm to determine recursively  the pairwise quantum discord in
a reduced density describing $k$ qubit system.

%%%%%%%%%%%%%%%%%%%%%%%%%%%%%%%%%%%%%%%%%%%%%%%%%%%%%%%%%%%%%%%%%%%%%%%%
\subsection{Two-qubit states }
%%%%%%%%%%%%%%%%%%%%%%%%%%%%%%%%%%%%%%%%%%%%%%%%%%%%%%%%%%%%%%%%%%%%%%%%
We begin with the two-qubit case. The tools we introduce are
useful when extending the size of the system to encompass more
qubits. We first consider the two-qubit states extracted from the
state (\ref{ncs}) by tracing out $(n-2)$ qubits. Since the $n$
qubits are all identical, we obtain $n(n-1)/2$ identical density
matrices. They are given by
\begin{equation}\label{rho12}
\rho_{12} = {\cal N}^2 \bigg[ \vert \eta, \eta \rangle \langle \eta, \eta \vert
+ e^{im\pi} q_2 \vert -\eta, -\eta \rangle \langle \eta, \eta  \vert$$
$$+ e^{-im\pi} q_2 \vert \eta, \eta  \rangle \langle -\eta, -\eta  \vert
+  \vert -\eta, -\eta  \rangle \langle -\eta, -\eta \vert
\bigg]
\end{equation}
where $q_2$ is defined by $q_s = p^{n-s}$ with $s=2$. The state (\ref{rho12}) can be alternatively written as
\begin{equation}
\rho_{12} = \frac{1}{2} ~(1 + p^{n-2})~ \frac{{\cal N}^2}{{{{\cal N}_2}_+}^2} ~\vert \eta \rangle_{2~2}\langle \eta \vert
 + \frac{1}{2} ~(1 - p^{n-2})~\frac{{\cal N}^2}{{{{\cal N}_2}_-}^2}~ Z \vert \eta \rangle_{2~2}\langle \eta \vert Z
\end{equation}
with
$$ \vert \eta \rangle_2 = {{\cal N}_2}_+ ( \vert \eta, \eta \rangle + e^{im\pi} \vert -\eta, -\eta \rangle) \quad  {\rm and} \quad
Z\vert \eta \rangle_2 = {{\cal N}_2}_- ( \vert \eta, \eta  \rangle - e^{im\pi} \vert -\eta, -\eta  \rangle). $$
The normalization factors  are defined by
$$ {{\cal N}_s}^{-2}_{\pm} = 2(1 \pm p^s\cos m\pi).$$
for $s=2$. In the computational base $\{ \vert 00 \rangle, \vert 01
\rangle, \vert 10 \rangle, \vert 11 \rangle \}$, the density matrix
$\rho_{12}$ has the form of the alphabet $X$. Indeed, it is
represented by
\begin{eqnarray}
\rho_{12} = 2{\cal N}^2 \left(
\begin{array}{cccc}
q_{2+}a^4_{+}
& 0 & 0 &  q_{2+}a^2_{+}a^2_{-} \\
0 & q_{2-}a^2_{+}a^2_{-} & q_{2-}a^2_{+}a^2_{-} & 0 \\
0 & q_{2-}a^2_{+}a^2_{-} & q_{2-}a^2_{+}a^2_{-}
& 0 \\
q_{2+}a^2_{+}a^2_{-} & 0 & 0 & q_{2+} a^4_{-}
\end{array}
\right) \,
\end{eqnarray}
where
$$ a_{\pm} = \frac{\sqrt{1\pm p}}{\sqrt{2}} \qquad {\rm and} \qquad q_{s\pm} = 1 \pm q_s \cos m\pi.$$
The state $\rho_{12}$ can be written also as
\begin{equation}\label{12-fano}
\rho_{12}= \sum_{k,l = 0,1} \rho^{kl} \otimes \vert k \rangle \langle l \vert.
\end{equation}
This form is suitable to establish a relation between the Bloch components of
the $2\times 2$ matrices $\rho^{kl}$ and the correlation matrix elements associated
with the two-qubit state $\rho_{12}$. In equation (\ref{12-fano}), the matrices $\rho^{ij}$
writes in Bloch representation as
\begin{equation}\label{0011}
 \rho^{00} =  \frac{1}{2} ( T^{00}_{0} \sigma_0 + T^{00}_{3} \sigma_3)\qquad  \rho^{11} = \frac{1}{2}  ( T^{11}_{0} \sigma_0 + T^{11}_{3} \sigma_3)
\end{equation}
and
\begin{equation}\label{0110}
 \rho^{01} = \frac{1}{2} ( T^{01}_{1} \sigma_1 + T^{01}_{2} \sigma_2 )  \qquad  \rho^{10} = \frac{1}{2} ( T^{10}_{1} \sigma_1 + T^{10}_{2} \sigma_2)
\end{equation}
where the Bloch components $T^{kl}_{\alpha}$ ($\alpha= 0, 1,2,3$) are
$$T^{kk}_{0} = {\cal N}^2 (1 + (-)^k p) (1 + (-)^k p^{n-1}\cos m\pi), \quad T^{kk}_{3} = {\cal N}^2 (1 + (-)^k p) (1 + (-)^k p^{n-2}\cos m\pi) $$
for $k = 0,1 $, and
$$ T^{01}_{1} = T^{10}_{1} = {\cal N}^2 (1 - p^2), \qquad T^{01}_{2} = - T^{10}_{2} = i {\cal N}^2 (1 - p^2)p^{n-2}\cos m\pi.$$
Reporting (\ref{0011}) and (\ref{0110}) in (\ref{12-fano}), one gets
\begin{equation}\label{rho12-sum}
\rho_{12} = \sum_{\alpha \beta} T_{\alpha \beta}
\sigma_{\alpha}\otimes \sigma_{\beta}
\end{equation}
where the non vanishing matrix elements $T_{\alpha \beta}$ $(\alpha,
\beta = 0,1,2,3)$ are given by
\begin{equation}\label{recurr}
T_{\alpha 0} = T^{00}_{\alpha} + T^{11}_{\alpha} ~~ {\rm for}~ \alpha = 0,3 \qquad  T_{\alpha 1} = T^{01}_{\alpha} + T^{10}_{\alpha}  ~~ {\rm for}~ \alpha = 1 $$
$$ T_{\alpha 2} = i T^{01}_{\alpha} -i T^{10}_{\alpha}  ~~ {\rm for}~ \alpha = 2 \qquad  T_{\alpha 3} = T^{00}_{\alpha} - T^{11}_{\alpha}  ~~ {\rm for}~ \alpha =0,3
\end{equation}
which gives
\begin{equation}\label{Tij12}
 T_{00} = 1, \quad T_{11} = 2{\cal N}^2 (1- p^2), \quad T_{22} = -2{\cal N}^2 (1- p^2)~p^{n-2}\cos
m\pi,$$ $$ T_{33} = 2{\cal N}^2 (p^2 + p^{n-2}\cos m\pi), \quad
T_{03} = T_{30} = 2{\cal N}^2 (p + p^{n-1}\cos m\pi).
\end{equation}
The expressions (\ref{recurr}) establish the relations between the  Bloch components  associated with one
qubit states (\ref{0011}) and (\ref{0110}) and the two qubit Fano-Bloch tensor elements  $T^{kl}_{\alpha}$ occurring in the two qubit density
$\rho_{12}$ (\ref{rho12-sum}). This result is generalizable to more qubits.  This issue is discussed in what follows.

%%%%%%%%%%%%%%%%%%%%%%%%%%%%%%%%%%%%%%%%%%%%%%%%%%%%%%%%%%%%%%%%%%%%%%%%%%%%%%%%%
\subsection{Three-qubit states}
%%%%%%%%%%%%%%%%%%%%%%%%%%%%%%%%%%%%%%%%%%%%%%%%%%%%%%%%%%%%%%%%%%%%%%%%%%%%%%%%%
The three-qubit states is extracted from the whole state (\ref{ncs})
by  removing $(n-3)$ qubits by the usual trace procedure. In this case, one obtains $n(n-1)(n-2)/3!$ density matrices which are all
identical. Explicitly,  they are given by
\begin{equation}\label{rho123}
\rho_{123} = {\cal N}^2 \bigg[ \vert \eta, \eta, \eta \rangle \langle \eta, \eta, \eta \vert
+ e^{im\pi} q_3 \vert -\eta, -\eta, -\eta \rangle \langle \eta, \eta, \eta \vert$$
$$+ e^{-im\pi} q_3 \vert \eta, \eta, \eta \rangle \langle -\eta, -\eta, -\eta \vert
+  \vert -\eta, -\eta, -\eta \rangle \langle -\eta, -\eta, -\eta \vert
\bigg]
\end{equation}
where $ q_3 = p^{n-3}$. Analogously to the previous case, we write the mixed three qubit state $\rho_{123}$  in a more compact form as follows
\begin{equation}
\rho_{123} = \frac{1}{2} ~(1 + p^{n-3})~ \frac{{\cal N}^2}{{\cal N}_{3+}^2} ~\vert \eta \rangle_{3~3}\langle \eta \vert
 + \frac{1}{2} ~(1 - p^{n-3})~\frac{{\cal N}^2}{{\cal N}_{3-}^2}~ Z \vert \eta \rangle_{3~3}\langle \eta \vert Z
\end{equation}
where
$$ \vert \eta \rangle_3 = {\cal N}_{3+} ( \vert \eta, \eta, \eta \rangle + e^{im\pi} \vert -\eta, -\eta, -\eta \rangle) \qquad
Z\vert \eta \rangle_3 = {\cal N}_{3-} ( \vert \eta, \eta, \eta \rangle - e^{im\pi} \vert -\eta, -\eta, -\eta \rangle) $$
with the normalization factors $ {\cal N}_{3\pm}$  given by
$$ {\cal N}^{-2}_{3\pm} = 2(1 \pm p^3\cos m\pi).$$
In the computational base $\{ \vert 000 \rangle, \vert 010 \rangle, \vert 100 \rangle, \vert 110 \rangle,
\vert 001 \rangle, \vert 011 \rangle, \vert101 \rangle, \vert 111 \rangle \}$, the state $\rho_{123}$
 takes the matrix form
 \begin{equation}\label{3X-class2}
\frac{\rho_{123}}{2{\cal N}^2}=\left(%
\begin{array}{cccccccc}
 q_{+3}a^6_{+} & 0 & 0 & q_{+3}a^4_{+}a^2_{-} & 0 & q_{+3}a^4_{+}a^2_{-}  & q_{+3}a^4_{+}a^2_{-}  & 0 \\
  0 &q_{-3}a^4_{+}a^2_{-} & q_{-3}a^4_{+}a^2_{-}  & 0 & q_{-3}a^4_{+}a^2_{-}  & 0 & 0 &q_{-3}a^2_{+}a^4_{-}  \\
  0 & q_{-3}a^4_{+}a^2_{-} & q_{-3}a^4_{+}a^2_{-}  & 0 & q_{-3}a^4_{+}a^2_{-}  & 0 & 0 &  q_{-3}a^2_{+}a^4_{-} \\
  q_{+3}a^4_{+}a^2_{-}  & 0 & 0 & q_{+3}a^2_{+}a^4_{-}  & 0 & q_{+3}a^2_{+}a^4_{-} & q_{+3}a^2_{+}a^4_{-} & 0 \\
  0 & q_{-3}a^4_{+}a^2_{-}  & q_{-3}a^4_{+}a^2_{-} & 0 &  q_{-3}a^4_{+}a^2_{-} & 0 & 0 &  q_{-3}a^2_{+}a^4_{-} \\
  q_{+3}a^4_{+}a^2_{-} & 0 & 0 & q_{+3}a^2_{+}a^4_{-} & 0 & q_{+3}a^2_{+}a^4_{-} & q_{+3}a^2_{+}a^4_{-} & 0 \\
q_{+3}a^4_{+}a^2_{-} & 0 & 0 & q_{+3}a^2_{+}a^4_{-} & 0 & q_{+3}a^2_{+}a^4_{-} & q_{+3}a^2_{+}a^4_{-} & 0 \\
  0 & q_{-3}a^2_{+}a^4_{-} & q_{-3}a^2_{+}a^4_{-} & 0 & q_{-3}a^2_{+}a^4_{-} & 0 & 0 & q_{-3}a^6_{-} \\
\end{array}%
\right)
\end{equation}
%{\bf Attention changer $q_{\pm}$ par $q_{\pm3}$}
The state (\ref{3X-class2}) can be also  re-written as
\begin{equation}\label{3X-fano1-calss2}
\rho_{123}= \sum_{k,l = 0,1} \rho^{kl} \otimes \vert k \rangle \langle l \vert
\end{equation}
where $\vert k \rangle , \vert l \rangle $ are related to the qubit $3$.
The two qubit density matrices $\rho^{kk}$ (for $k = 0,1$) writes, in the computational basis spanned by  $\{  |0 \rangle_1 \otimes |0\rangle_2,
|0\rangle_1 \otimes |1\rangle_2 , |1\rangle_1 \otimes
|0\rangle_2,  |1\rangle_1 \otimes |1\rangle_2\} $, as
\begin{eqnarray}
\rho^{00}= 2{\cal N}^2\left(
\begin{array}{cccc}
q_{+3}a^6_{+}
& 0 & 0 &  q_{+3}a^4_{+}a^2_{-} \\
0 & q_{-3}a^4_{+}a^2_{-} & q_{-3}a^4_{+}a^2_{-} & 0 \\
0 & q_{-3}a^4_{+}a^2_{-} & q_{-3}a^4_{+}a^2_{-}
& 0 \\
q_{+3}a^4_{+}a^2_{-} & 0 & 0 & q_{+3}a^2_{+}a^4_{-}
\end{array}
\right) \,,\label{2X-00}
\end{eqnarray}
and
\begin{eqnarray}
\rho^{11}= 2{\cal N}^2\left(
\begin{array}{cccc}
 q_{-3}a^4_{+}a^2_{-}
& 0 & 0 &  q_{-3}a^2_{+}a^4_{-} \\
0 & q_{+3}a^2_{+}a^4_{-} & q_{+3}a^2_{+}a^4_{-} & 0 \\
0 & q_{+3}a^2_{+}a^4_{-} & q_{+3}a^2_{+}a^4_{-}
& 0 \\
q_{-3}a^2_{+}a^4_{-} & 0 & 0 & q_{-3}a^6_{-}
\end{array}
\right) \,. \label{2X-11}
\end{eqnarray}
For $(k = 0 , l = 1)$ and $(k = 1 , l = 0)$ , we have respectively
\begin{eqnarray}
\rho^{01} = 2{\cal N}^2 \left(
\begin{array}{cccc}
0
& q_{+3}a^4_{+}a^2_{-} & q_{+3}a^4_{+}a^2_{-}  & 0 \\
q_{-3}a^4_{+}a^2_{-} & 0 & 0  & q_{-3}a^2_{+}a^4_{-} \\
q_{-3}a^4_{+}a^2_{-} & 0 & 0
& q_{-3}a^2_{+}a^4_{-}\\
0 & q_{+3}a^2_{+}a^4_{-} & q_{+3}a^2_{+}a^4_{-} & 0
\end{array}
\right) \,,  \label{2X-01}
\end{eqnarray}
and
\begin{eqnarray}
\rho^{10} = 2{\cal N}^2 \left(
\begin{array}{cccc}
0
& q_{-3}a^4_{+}a^2_{-} & q_{-3}a^4_{+}a^2_{-} & 0 \\
 q_{+3}a^4_{+}a^2_{-} & 0 & 0  & q_{+3}a^2_{+}a^4_{-} \\
 q_{+3}a^4_{+}a^2_{-} & 0 & 0
& q_{+3}a^2_{+}a^4_{-}\\
0 &  q_{-3}a^2_{+}a^4_{-} &  q_{-3}a^2_{+}a^4_{-} & 0
\end{array}
\right) \, . \label{2X-10}
\end{eqnarray}
The Fano-Bloch representation of the matrices $\rho^{kk}$, given by
(\ref{2X-00}) and (\ref{2X-11}), take the form
\begin{equation}\label{2X-fano-class2-1}
\rho^{kk}=\frac{1}{4}  \sum_{\alpha \beta} T^{kk}_{\alpha \beta} ~ \sigma_{\alpha}\otimes \sigma_{\beta}
\end{equation}
where $\alpha, \beta = 0,1,2,3$ and the correlation matrix elements $T^{kk}_{\alpha \beta}$ are given by
$$T^{kk}_{\alpha \beta} = {\rm Tr}(\rho^{kk}~\sigma_{\alpha}\otimes \sigma_{\beta}).$$
The explicit expressions of the non vanishing contributions are
\begin{equation}\label{Tii}
T^{kk}_{00}= 1$$
$$ T^{kk}_{30}= T^{kk}_{03} = \frac{p}{2}~ (1 + (-)^kp)~\frac{1 + (-)^kp^{n-3}\cos m\pi}{1 + p^{n}\cos m\pi}$$
$$T^{kk}_{11}=\frac{1}{2}  ~(1 + (-)^kp)~\frac{1 - p^{2}}{1 + p^{n}\cos m\pi}$$
$$T^{kk}_{22}= - \frac{1}{2} ~ (1 + (-)^kp)~\frac{(1 - p^{2}) p^{n-3}\cos m\pi}{1 + p^{n}\cos m\pi}$$
$$T^{kk}_{33}=\frac{1}{2} ~ (1 + (-)^kp)~\frac{p^2 + (-)^kp^{n-3}\cos m\pi}{1 + p^{n}\cos m\pi}
\end{equation}
Similarly, for the two-qubit states $\rho^{kl}$ ($ k \neq l$) given by (\ref{2X-01}) and (\ref{2X-10}), the Fano-Bloch representation writes
\begin{equation}\label{2X-fano-class2-2}
\rho^{kl}=\frac{1}{4}  \sum_{\alpha \beta} T^{kl}_{\alpha \beta} ~ \sigma_{\alpha}\otimes \sigma_{\beta}
\end{equation}
where the non zero matrix elements $T^{kl}_{\alpha \beta}$ are given by
\begin{equation}\label{Tij}
T^{kl}_{01}= T^{kl}_{10}= \frac{1}{2}~ \frac{1-p^2}{1 + p^{n}\cos m\pi} $$
$$T^{kl}_{02}= T^{kl}_{20}=  (-)^k \frac{i}{2}~ \frac{p(1-p^2)}{1 + p^{n}\cos m\pi}  $$
$$T^{kl}_{13}= T^{kl}_{31}= \frac{1}{2}~ \frac{(1-p^2)p^{n-2}\cos m\pi)}{1 + p^{n}\cos m\pi} $$
$$T^{kl}_{23}= T^{kl}_{32}= (-)^k \frac{i}{2}~ \frac{(1-p^2)p^{n-2}\cos m\pi)}{1 + p^{n}\cos m\pi}.
\end{equation}
Using  (\ref{3X-fano1-calss2}), the three-qubit state $\rho_{123}$ expands  as
\begin{equation}\label{3X-fano1}
\rho_{123}=\frac{1}{2} \Bigg[ (\rho^{00} + \rho^{11} )\otimes \sigma_0  +   (\rho^{00} - \rho^{11} )\otimes \sigma_3 +
 (\rho^{01} + \rho^{10} )\otimes \sigma_1  + i (\rho^{01} - \rho^{10} )\otimes \sigma_2 \Bigg]
\end{equation}
Inserting  (\ref{2X-fano-class2-1}) and  (\ref{2X-fano-class2-2}) in the equation (\ref{3X-fano1}) and using  the results (\ref{Tii}) and (\ref{Tij}) , one gets
\begin{equation}\label{density}
\rho_{123} = \frac{1}{8} \sum_{\alpha\beta} \Bigg[T_{\alpha\beta0} ~\sigma_{\alpha} \otimes \sigma_{\beta} \otimes \sigma_{0}+
T_{\alpha\beta1} ~\sigma_{\alpha} \otimes \sigma_{\beta} \otimes \sigma_{1}+
T_{\alpha\beta2} ~\sigma_{\alpha} \otimes \sigma_{\beta} \otimes \sigma_{2}+
T_{\alpha\beta3} ~\sigma_{\alpha} \otimes \sigma_{\beta} \otimes \sigma_{3}\Bigg]
\end{equation}
where
\begin{equation}\label{relation3ret2r-class2}
T_{\alpha\beta0}=    T^{++}_{\alpha\beta} = T^{00}_{\alpha\beta} + T^{11}_{\alpha\beta} $$
$$ T_{\alpha\beta3}= T^{--}_{\alpha\beta} = T^{00}_{\alpha\beta} - T^{11}_{\alpha\beta}
\end{equation}
with $\alpha \beta = 00, 03, 30,  11, 22, 33 $ (cf. (\ref{Tii})),
%$$ (00), (03), (30),  (11), (22), (33)$$
and
\begin{equation}\label{relation3ret2r-class22}
T_{\alpha\beta1} = T^{+-}_{\alpha\beta} = T^{01}_{\alpha\beta} + T^{10}_{\alpha\beta}$$
$$ T_{\alpha\beta2} = T^{-+}_{\alpha\beta} = iT^{01}_{\alpha\beta} - iT^{10}_{\alpha\beta}.
\end{equation}
with $\alpha \beta= 01, 02, 10, 20 , 13 , 23, 31, 32$ (cf.
(\ref{Tij})). Reporting (\ref{Tii}) and (\ref{Tij}) in the
expressions (\ref{relation3ret2r-class2}) and
(\ref{relation3ret2r-class22}), one obtains the 32 non vanishing
correlation matrix elements  $T_{\alpha\beta\gamma}$ corresponding
to the three qubit state $\rho_{123}$. Subsequently, the recursive
relations (\ref{relation3ret2r-class2}) and
(\ref{relation3ret2r-class22}) offer a nice tool to determine the
correlation elements $T_{\alpha\beta\gamma}$ in terms of those
associated with the two qubit density
matrices $\rho^{kk}$ and $\rho^{kl}$ given, respectively,  by (\ref{2X-fano-class2-1}) and (\ref{2X-fano-class2-2}). %It must be
%noticed that the recursive relations (\ref{relation3ret2r-class2}) and (\ref{relation3ret2r-class22}) offers a key ingredient to determine
%the correlations coefficients occurring in the Fano-Bloch representation of the density matrix $\rho_{123}$ (\ref{density}) and those associated with
%the sub-matrices $\rho^{kk}$ and $\rho^{kl}$ given, respectively,  by (\ref{2X-fano-class2-1}) and (\ref{2X-fano-class2-2}).
 Clearly, along the same line of reasoning, the recursive  relation obtained for two and three qubits  are ready to be extended  to an arbitrary $k$ qubit state.

\subsection{$k$-qubit states}
%%%%%%%%%%%%%%%%%%%%%%%%%%%%%%%%%%%%%%%%%%%%%%%%%%%%%%%%%%%%%%%%%%%%%%%%
A mixed $k$-qubit state $(k= 2, 3, \cdots, n)$ is obtained by tracing
out $(n-k)$ qubit from the state (\ref{ncs}). It is given by
\begin{equation}\label{rho123k-trace}
\rho_{123\cdots k} = {\cal N}^2 \bigg[ \vert \eta, \eta,\cdots, \eta \rangle \langle \eta, \eta,\cdots, \eta \vert
+ e^{im\pi} q_k \vert -\eta, -\eta,\cdots, -\eta \rangle \langle \eta, \eta,\cdots, \eta \vert$$
$$+ e^{-im\pi} q_k \vert \eta, \eta,\cdots, \eta \rangle \langle -\eta, -\eta,\cdots, -\eta \vert
+  \vert -\eta, -\eta,\cdots, -\eta \rangle \langle -\eta, -\eta,\cdots -\eta \vert
\bigg]
\end{equation}
where $ q_k = p^{n-k}$. The reduced density matrix $\rho_{123\cdots k}$ is of rank 2. Indeed, the state
(\ref{rho123k-trace}) rewrites
\begin{equation}
\rho_{123\cdots k} = \frac{1}{2} ~(1 + p^{n-k})~ \frac{{\cal N}^2}{{\cal N}_{k+}^2} ~\vert \eta \rangle_{k~k}\langle \eta \vert
 + \frac{1}{2} ~(1 - p^{n-k})~\frac{{\cal N}^2}{{\cal N}_{k-}^2}~ Z \vert \eta \rangle_{k~k}\langle \eta \vert Z
\end{equation}
where
$$ \vert \eta \rangle_k = {\cal N}_{k+} ( \vert \eta, \eta,\cdots, \eta \rangle + e^{im\pi} \vert -\eta, -\eta,\cdots, -\eta \rangle) \qquad
Z\vert \eta \rangle_k = {\cal N}_{k-} ( \vert \eta, \eta,\cdots, \eta \rangle - e^{im\pi} \vert -\eta, -\eta,\cdots, -\eta \rangle) $$
and the normalization factors $ {\cal N}_{k\pm}$ are given by
$$ {\cal N}^{-2}_{k\pm} = 2(1 \pm p^k\cos m\pi).$$
The cyclic operator $Z$ is now defined by
$$ Z \vert \eta, \eta, \cdots, \eta \rangle = \vert \eta, \eta, \cdots ,\eta \rangle \qquad  Z \vert - \eta, - \eta, \cdots , -\eta \rangle = -  \vert - \eta, - \eta, \cdots , -\eta \rangle.$$
Using (\ref{rho123k-trace}), it is simple to check that the $k$-qubit state $\rho_{123\cdots k}$ can
be expressed in terms of states comprising $(k-1)$-qubits. The state $\rho_{123\cdots k}$ (\ref{rho123k-trace})  can be written also as
\begin{equation}\label{rho123k-decompose}
\rho_{123\cdots k} = \sum_{rs=1,2} \rho^{rs}_{12\cdots (k-1)} \otimes \vert r \rangle \langle s \vert
\end{equation}
where
\begin{equation}
\rho^{rs}_{12\cdots (k-1)} \equiv \rho^{rs} =  a_+^{2-r-s}a_-^{r+s} \Bigg[ \frac{1}{2} ~(1 + p^{n-k})~
\frac{{\cal N}^2}{{\cal N}_{(k-1)+}^2} ~ Z^r\vert \eta \rangle_{(k-1)~(k-1)}\langle \eta \vert Z^s$$
$$ + \frac{1}{2} ~(1 - p^{n-k})~\frac{{\cal N}^2}{{\cal N}_{(k-1)-}^2}~ Z^{r+1} \vert \eta \rangle_{(k-1)~(k-1)}\langle \eta \vert Z^{s+1}\Bigg]
\end{equation}
Explicitly, the $k$-qubit matrix  (\ref{rho123k-decompose}) writes
\begin{equation}\label{rho123k}
\rho_{123\cdots k} = \frac{1}{2} (\rho^{00} + \rho^{11}) \otimes \sigma_{0} +
\frac{1}{2} (\rho^{01}+ \rho^{10}) \otimes \sigma_{1} + \frac{i}{2} (\rho^{01} - \rho^{10}) \otimes \sigma_{2}
+ \frac{1}{2} (\rho^{00} - \rho^{11}) \otimes \sigma_{3}
\end{equation}
and the $(k-1)$-qubit states $\rho^{rs}$ can be expanded, in Fano-Bloch representation,  as
\begin{equation}\label{fanorhors}
\rho^{rs}= \frac{1}{2^{k-1}} \sum_{\alpha_1,\alpha_2, \cdots,\alpha_{k-1}} T^{rs}_{\alpha_1 \alpha_2 \cdots \alpha_{k-1}}
\sigma_{\alpha_1} \otimes \sigma_{\alpha_2} \otimes \cdots \otimes \sigma_{\alpha_{k-1}}.
\end{equation}
Hence, reporting (\ref{fanorhors}) in (\ref{rho123k-decompose}), the $k$-qubit state $\rho_{123\cdots k}$ takes the form
\begin{equation}\label{fanorhor123k}
\rho_{123\cdots k} = \frac{1}{2^{k}} \sum_{\alpha_1,\alpha_2, \cdots,\alpha_{k-1},\alpha_{k}} T_{\alpha_1 \alpha_2 \cdots \alpha_{k-1}\alpha_k}
\sigma_{\alpha_1} \otimes \sigma_{\alpha_2} \otimes \cdots \otimes \sigma_{\alpha_{k-1}}\otimes \sigma_{\alpha_{k}}.
\end{equation}
where the correlation matrix elements $T_{\alpha_1 \alpha_2 \cdots \alpha_{k-1}\alpha_k}$ express in terms of the correlations
coefficients occurring in (\ref{fanorhors}) as
\begin{equation}\label{recur-k}
  T_{\alpha_1 \alpha_2 \cdots \alpha_{k-1}0} = T^{00}_{\alpha_1 \alpha_2 \cdots \alpha_{k-1}} + T^{11}_{\alpha_1 \alpha_2 \cdots \alpha_{k-1}} $$
$$ T_{\alpha_1 \alpha_2 \cdots \alpha_{k-1}3}= T^{00}_{\alpha_1 \alpha_2 \cdots \alpha_{k-1}} - T^{11}_{\alpha_1 \alpha_2 \cdots \alpha_{k-1}}$$
$$ T_{\alpha_1 \alpha_2 \cdots \alpha_{k-1}1} = T^{01}_{\alpha_1 \alpha_2 \cdots \alpha_{k-1}} + T^{10}_{\alpha_1 \alpha_2 \cdots \alpha_{k-1}}$$
$$ T_{\alpha_1 \alpha_2 \cdots \alpha_{k-1}2} = i T^{01}_{\alpha_1 \alpha_2 \cdots \alpha_{k-1}} -i T^{10}_{\alpha_1 \alpha_2 \cdots \alpha_{k-1}},
\end{equation}
and we have the relations between the correlation matrix elements of $k$ and $k-1$-qubit states. In this picture the correlation matrix elements
associated with a  $k$-qubit state can be recursively expressed in terms of ones involving two qubits. It is simply verified that the relations (\ref{recur-k}) reduce to (\ref{recurr}) for
$k=2$ and to (\ref{relation3ret2r-class2}-\ref{relation3ret2r-class22}) for $k=3$. To illustrate the algorithm in deriving relations of type (\ref{recur-k}),
 we consider the case of four qubits. In this situation, the density matrix (\ref{rho123k-trace}) becomes
\begin{equation}\label{rho1234}
\rho_{1234} = {\cal N}^2 \bigg[ \vert \eta, \eta,\eta, \eta \rangle \langle \eta, \eta,\eta, \eta \vert
+ e^{im\pi} q_4 \vert -\eta, -\eta,-\eta, -\eta \rangle \langle \eta, \eta,\eta, \eta \vert$$
$$+ e^{-im\pi} q_4 \vert \eta, \eta,\eta, \eta \rangle \langle -\eta, -\eta,-\eta, -\eta \vert
+  \vert -\eta, -\eta,-\eta, -\eta \rangle \langle -\eta, -\eta,-\eta -\eta \vert
\bigg]
\end{equation}
and the expression (\ref{rho123k-decompose}) gives
\begin{equation}\label{rho1234-bis}
\rho_{1234} = \rho_{123}^{00}\otimes \vert 0\rangle \langle 0\vert + \rho_{123}^{01}\otimes \vert0 \rangle \langle 1\vert + \rho_{123}^{10}\otimes \vert 1\rangle \langle 0 \vert
+ \rho_{123}^{11}\otimes \vert 1\rangle \langle 1 \vert
\end{equation}
where the three-qubit states $\rho_{123}^{00}$, $\rho_{123}^{01}$, $\rho_{123}^{10}$ and $\rho_{123}^{11}$ are given in the
usual computational basis as
\begin{equation}\label{rho00-4}
\frac{\rho^{00}_{123}}{2{\cal N}^2}=\left(%
\begin{array}{cccccccc}
 q_{+4}a^8_{+} & 0 & 0 & q_{+4}a^6_{+}a^2_{-} & 0 & q_{+4}a^6_{+}a^2_{-}  & q_{+4}a^6_{+}a^2_{-}  & 0 \\
  0 &q_{-4}a^6_{+}a^2_{-} & q_{-4}a^6_{+}a^2_{-}  & 0 & q_{-4}a^6_{+}a^2_{-}  & 0 & 0 &q_{-4}a^4_{+}a^4_{-}  \\
  0 & q_{-4}a^6_{+}a^2_{-} & q_{-4}a^6_{+}a^2_{-}  & 0 & q_{-4}a^6_{+}a^2_{-}  & 0 & 0 &  q_{-4}a^4_{+}a^4_{-} \\
  q_{+4}a^6_{+}a^2_{-}  & 0 & 0 & q_{+4}a^4_{+}a^4_{-}  & 0 & q_{+4}a^4_{+}a^4_{-} & q_{+4}a^4_{+}a^4_{-} & 0 \\
  0 & q_{-4}a^6_{+}a^2_{-}  & q_{-4}a^6_{+}a^2_{-} & 0 &  q_{-4}a^6_{+}a^2_{-} & 0 & 0 &  q_{-4}a^4_{+}a^4_{-} \\
  q_{+4}a^6_{+}a^2_{-} & 0 & 0 & q_{+4}a^4_{+}a^4_{-} & 0 & q_{+4}a^4_{+}a^4_{-} & q_{+4}a^4_{+}a^4_{-} & 0 \\
q_{+4}a^6_{+}a^2_{-} & 0 & 0 & q_{+4}a^4_{+}a^4_{-} & 0 & q_{+4}a^4_{+}a^4_{-} & q_{+4}a^4_{+}a^4_{-} & 0 \\
  0 & q_{-4}a^4_{+}a^4_{-} & q_{-4}a^4_{+}a^4_{-} & 0 & q_{-4}a^4_{+}a^4_{-} & 0 & 0 & q_{-4}a^2_{+}a^6_{-} \\
\end{array}%
\right)
\end{equation}
\begin{equation}\label{rho11-4}
\frac{\rho^{11}_{123}}{2{\cal N}^2}=\left(%
\begin{array}{cccccccc}
 q_{-4}a^6_{+}a^2_{-} & 0 & 0 & q_{-4}a^4_{+}a^4_{-} & 0 & q_{-4}a^4_{+}a^4_{-}  & q_{-4}a^4_{+}a^4_{-}  & 0 \\
  0 &q_{+4}a^4_{+}a^4_{-} & q_{+4}a^4_{+}a^4_{-}  & 0 & q_{+4}a^4_{+}a^4_{-}  & 0 & 0 &q_{+4}a^2_{+}a^6_{-}  \\
  0 & q_{+4}a^4_{+}a^4_{-} & q_{+4}a^4_{+}a^4_{-}  & 0 & q_{+4}a^4_{+}a^4_{-}  & 0 & 0 &  q_{+4}a^2_{+}a^6_{-} \\
  q_{-4}a^4_{+}a^4_{-}  & 0 & 0 & q_{-4}a^2_{+}a^6_{-}  & 0 & q_{-4}a^2_{+}a^6_{-} & q_{-4}a^2_{+}a^6_{-} & 0 \\
  0 & q_{+4}a^4_{+}a^4_{-}  & q_{+4}a^4_{+}a^4_{-} & 0 &  q_{+4}a^4_{+}a^4_{-} & 0 & 0 &  q_{+4}a^2_{+}a^6_{-} \\
  q_{-4}a^4_{+}a^4_{-} & 0 & 0 & q_{-4}a^2_{+}a^6_{-} & 0 & q_{-4}a^2_{+}a^6_{-} & q_{-4}a^2_{+}a^6_{-} & 0 \\
q_{-4}a^4_{+}a^4_{-} & 0 & 0 & q_{-4}a^2_{+}a^6_{-} & 0 & q_{-4}a^2_{+}a^6_{-} & q_{-4}a^2_{+}a^6_{-} & 0 \\
  0 & q_{+4}a^2_{+}a^6_{-} & q_{+4}a^2_{+}a^6_{-} & 0 & q_{+4}a^2_{+}a^6_{-} & 0 & 0 & q_{+4} a^8_{-} \\
\end{array}%
\right),
\end{equation}
\begin{equation}\label{rho01-4}
\frac{\rho^{01}_{123}}{2{\cal N}^2}=\left(%
\begin{array}{cccccccc}
 0 & q_{+4}a^6_{+}a^2_{-} &  q_{+4}a^6_{+}a^2_{-} & 0 &  q_{-4}a^6_{+}a^2_{-} & 0  & 0  & q_{-4}a^4_{+}a^4_{-} \\
q_{-4}a^6_{+}a^2_{-} & 0 & 0  & q_{-4}a^4_{+}a^4_{-} & 0   & q_{+4}a^4_{+}a^4_{-}  & q_{+4}a^4_{+}a^4_{-} & 0  \\
q_{-4}a^6_{+}a^2_{-} & 0 & 0 & q_{-4}a^4_{+}a^4_{-} & 0   &  q_{+4}a^4_{+}a^4_{-} & q_{+4}a^4_{+}a^4_{-} & 0 \\
0 & q_{+4}a^4_{+}a^4_{-} & q_{+4}a^4_{+}a^4_{-} & 0  &  q_{-4}a^4_{+}a^4_{-} & 0 & 0 & q_{-4}a^2_{+}a^6_{-} \\
q_{-4}a^6_{+}a^2_{-} & 0  & 0 &  q_{-4}a^4_{+}a^4_{-} & 0 &   q_{-4}a^4_{+}a^4_{-} &  q_{-4}a^4_{+}a^4_{-} &  0 \\
0 &  q_{+4}a^4_{+}a^4_{-} &  q_{+4}a^4_{+}a^4_{-} & 0 &  q_{+4}a^4_{+}a^4_{-} & 0 & 0 &  q_{+4}a^2_{+}a^6_{-}\\
0 &  q_{+4}a^4_{+}a^4_{-} & q_{+4}a^4_{+}a^4_{-} & 0 & q_{+4}a^4_{+}a^4_{-} & 0 & 0 & q_{+4}a^2_{+}a^6_{-}\\
q_{-4}a^4_{+}a^4_{-} & 0 & 0 & q_{-4}a^2_{+}a^6_{-} & 0 & q_{-4}a^2_{+}a^6_{-} & q_{-4}a^2_{+}a^6_{-}& 0 \\
\end{array}%
\right),
\end{equation}
and
\begin{equation}\label{rho01-4}
\rho^{10}_{123} = \big( \rho^{01}_{123} \big)^t
\end{equation}
It is clear that with increasing the qubits number,  complicated analytical computation emerges especially
in computing the quantum correlations.  However, the recursive algorithm presented above, offers
an alternative way for symmetric multi-qubit (\ref{ncs}), to reduce the complexity in determining
analytical evaluation of geometric discord.  The expression (\ref{rho1234-bis}) allows us to express the correlations
factors $T_{\alpha_1 \alpha_2 \alpha_3 \alpha_4}$ in terms
of those corresponding to three-qubit density matrices $\rho^{00}_{123}$, $\rho^{01}_{123}$, $\rho^{10}_{123}$
and $\rho^{11}_{123}$. Indeed, the state $\rho_{1234}$ writes in the Fano-Bloch representation as
\begin{equation}\label{rho123k-bis}
\rho_{1234} = \frac{1}{2^4}\sum_{\alpha_1, \alpha_2, \alpha_3, \alpha_4} T_{\alpha_1 \alpha_2 \alpha_3 \alpha_4} \sigma_{\alpha_1}\otimes \sigma_{\alpha_2}
\otimes \sigma_{\alpha_3} \otimes \sigma_{\alpha_4}
\end{equation}
and by re-equating (\ref{rho1234-bis}) as
\begin{equation}\label{rho1234-bis-bis}
\rho_{1234} = \frac{1}{2} (\rho_{123}^{00}+ \rho_{123}^{11}) \otimes \sigma_0 + \frac{1}{2} (\rho_{123}^{01}+ \rho_{123}^{10})\otimes \sigma_1
+ \frac{i}{2} (\rho_{123}^{01} - \rho_{123}^{10}) \otimes \sigma_2
+ \frac{1}{2} (\rho_{123}^{00} - \rho_{123}^{00}) \otimes \sigma_3,
\end{equation}
it is simple to see that
\begin{equation}\label{recu-4}
T_{\alpha_1 \alpha_2 \alpha_3 0}  = T^{00}_{\alpha_1 \alpha_2 \alpha_3} + T^{11}_{\alpha_1 \alpha_2 \alpha_3} $$
$$ T_{\alpha_1 \alpha_2 \alpha_3 1} = T^{01}_{\alpha_1 \alpha_2 \alpha_3} + T^{10}_{\alpha_1 \alpha_2 \alpha_3}$$
$$ T_{\alpha_1 \alpha_2 \alpha_3 2} = iT^{01}_{\alpha_1 \alpha_2 \alpha_3} - i T^{10}_{\alpha_1 \alpha_2 \alpha_3}$$
$$ T_{\alpha_1 \alpha_2 \alpha_3 3}= T^{00}_{\alpha_1 \alpha_2 \alpha_3} - T^{11}_{\alpha_1 \alpha_2 \alpha_3}
\end{equation}
where the quantities $T^{kl}_{\alpha_1, \alpha_2, \alpha_3}$, defined so that
\begin{equation}\label{rhokl123}
\rho^{kl}_{123} = \frac{1}{2^3}\sum_{\alpha_1, \alpha_2, \alpha_3} T^{kl}_{\alpha_1 \alpha_2 \alpha_3} \sigma_{\alpha_1}\otimes \sigma_{\alpha_2}
\otimes \sigma_{\alpha_3},
\end{equation}
can be obtained easily following the method developed above for three and two qubit states. It
follows that the non vanishing elements $T_{\alpha_1 \alpha_2 \alpha_3 \alpha_4}$  are
those with indices  $({\alpha_1, \alpha_2, \alpha_3, \alpha_4})$ belonging to the following  set of quadruples
$$\{ \{ 00, 11, 22, 33, 03, 30\} \times \{0,3\} \times \{ 0,3\},$$
$$ \{ 00, 11, 22, 33, 03, 30\} \times \{1,2\} \times \{ 1,2\},$$
$$ \{ 01, 10, 20, 02, 13, 31, 23, 32\} \times \{1,2\} \times \{ 0,3\},$$
$$ \{ 01, 10, 20, 02, 13, 31, 23, 32\} \times \{0,3\} \times \{ 1,2\} \}.$$

Finally, we stress the usefulness of the recursive approach,
discussed in this section,  in determining the Fano-Bloch components
for an arbitrary $k$-qubit state in terms of those involving
$(k-1)$-qubits. This gives a simple way to specify the correlation
matrix elements for $k$-qubits state  in terms of ones associated
with two qubit subsystems. In this picture, for the symmetric
multi-qubit states (\ref{ncs}), considerable simplification arises
in establishing such recursive relations and subsequently simplify
drastically the evaluation of pairwise geometric quantum discord.

%%%%%%%%%%%%%%%%%%%%%%%%%%%%%%%%%%%%%%%%%%%%%%%%%%%%%%%%%%%%%%%%%%%%%%%%
\section{Geometric measure of quantum discord and classical states}
%%%%%%%%%%%%%%%%%%%%%%%%%%%%%%%%%%%%%%%%%%%%%%%%%%%%%%%%%%%%%%%%%%%%%%%%

We now face the question of determining the explicit form of
geometric discord between a qubit and a second party of dimension
$2^{k-1}$ in  the $k$-qubit mixed state (\ref{rho123k-trace}).  For
this end, we must first find the expression of closest classical
states to the states of type (\ref{rho123k-trace}) when the distance
is measured by Hilbert-Schmidt trace.  We shall follow the procedure
developed in \cite{Dakic2010} for a two qubit system.

%%%%%%%%%%%%%%%%%%%%%%%%%%%%%%%%%%%%%%%%%%%%%%%%%%%%%%%%%%%%%%%%%%%%%%%%
\subsection{Two-qubit states}
%%%%%%%%%%%%%%%%%%%%%%%%%%%%%%%%%%%%%%%%%%%%%%%%%%%%%%%%%%%%%%%%%%%%%%%%

For the two-qubit state (\ref{rho12-sum}) which rewrites

\begin{equation}\label{rho12sum}
\rho_{12} = \frac{1}{4}\Bigg[ \sigma_{0}\otimes \sigma_{0} +  T_{30}~
\sigma_{3}\otimes \sigma_{0}+ T_{03}~
\sigma_{0}\otimes \sigma_{3} +T_{11}~
\sigma_{1}\otimes \sigma_{1} + T_{22}~
\sigma_{2}\otimes \sigma_{2}+T_{33}~
\sigma_{3}\otimes \sigma_{3}\Bigg],
\end{equation}
 the zero-discord or classical states are given by
\begin{equation}
\chi_{12} = p_1 \vert \psi_1 \rangle \langle \psi_1 \vert \otimes \rho_1^{2} + p_2 \vert \psi_2 \rangle \langle \psi_2 \vert \otimes \rho_2^{2}
\end{equation}
where $\{ \vert \psi_1 \rangle , \vert \psi_2 \rangle \}$ is an orthonormal basis related to the qubit $1$ and $\rho_i^{2}$ $(i = 1, 2)$ are
reduced density matrices attached  the second qubit. It can be
written also as
\begin{equation}\label{xi12}
\chi_{12} = \frac{1}{4}\Bigg[ \sigma_{0}\otimes \sigma_{0}
+ \sum_{i=1}^3 te_i ~~\sigma_i \otimes \sigma_{0} + \sum_{i=1}^3 (s_+)_{i} ~\sigma_0 \otimes \sigma_{i}
+ \sum_{i,j=1}^3  e_i (s_-)_{j} ~\sigma_i \otimes \sigma_{j}\Bigg]
\end{equation}
where
$$ t = p_1-p_2 \qquad e_i = \langle \psi_1 \vert \sigma_i \vert \psi_1 \rangle  \qquad
(s_{\pm})_{j} = {\rm Tr}\big((p_1\rho_1^{2} \pm p_2\rho_2^{2}) \sigma_{j} \big)$$
%Remark that the component of the vector $\vec{e}=(e_1,e_2,e_3)^t$ satisfy $e_1^2 + e_2^2 + e_3^2 = 1$.
The distance between the density matrix $\rho_{12}$ (\ref{rho12sum}) and the classical state $\chi_{12}$ (\ref{xi12}), as
measured by Hilbert-Schmidt norm, is
\begin{equation}\label{HS12}
|| \rho_{12} - \chi_{12}||^2 = \frac{1}{4} \bigg[  (t^2 -2te_3 T_{30} + T^2_{30}) +  \sum_{i=1}^{3}
(T_{0i} -  (s_+)_{i})^2 + \sum_{i,j=1}^3   ( T_{ij} -  e_i(s_-)_{j})^2  \bigg]
\end{equation}
%To get the closest classical state to the quantum state $\rho_{12}$, one performs the
The minimal distance is obtained by minimizing  the the Hilbert-Schmidt norm (\ref{HS12}) with respect to the parameters $t$, $(s_+)_{i}$ and $(s_-)_{i}$. This gives
\begin{equation}\label{minHS12}
t= e_3T_{30} $$
$$ (s_+)_{1}= 0 \quad (s_+)_{2} = 0 \quad  (s_+)_{3} = T_{03} $$
$$ (s_-)_{i} = \sum_{j=1}^3 e_j T_{ji}
%$$  (s_-)_{1} = e_1 T_{11} \quad (s_-)_{2} = e_2 T_{22} \quad (s_-)_{3} = e_3 T_{33}
\end{equation}
Inserting  the solutions (\ref{minHS12}) in (\ref{HS12}), one gets
\begin{equation}\label{HS12-compact}
|| \rho_{12} - \chi_{12}||^2 = \frac{1}{4} \bigg[  {\rm Tr}K - \vec{e}^t K \vec{e} \bigg]
\end{equation}
where the matrix $K$ is defined by
\begin{equation}\label{matrixK12}
K = {\rm diag}(T_{11}^2, T_{22}^2, T_{30}^2+T_{33}^2)
\end{equation}
From  (\ref{Tij12}), the eigenvalues of the  matrix $K$ (\ref{matrixK12}) read
\begin{equation}\label{GD12lambda1}
\lambda_1 = \frac{(1-p^2)^2}{(1+p^n\cos m\pi)^2}
\end{equation}
\begin{equation}\label{GD12lambda2}
\lambda_2 = \frac{(1-p^2)^2 p^{2(n-2)}}{(1+p^n\cos m\pi)^2}
\end{equation}
\begin{equation}\label{GD12lambda3}
\lambda_3 = \frac{(p^2+ p^{2(n-2)})(1+p^2) + 4 p^n\cos m\pi}{(1+p^n\cos m\pi)^2}
\end{equation}
It easily seen from (\ref{HS12-compact}) that  the minimal Hilbert-Schmidt distance is obtained for the vector $\vec{e}$ associated with the maximal eigenvalue $\lambda_{\rm max}$
of the matrix $K$. Thus, the geometric measure of quantum discord in the state $\rho_{12}$ is  given by
\begin{equation}\label{GD12}
D_{\rm g}(\rho_{12}) = \frac{1}{4} (\lambda_{1}+\lambda_{2}+\lambda_{3}-\lambda_{\rm max})
\end{equation}
From the expressions (\ref{GD12lambda1}) and (\ref{GD12lambda2}), we
have $\lambda_2 < \lambda_1$. This implies that $\lambda_{\rm max}$
is equal to $\lambda_1$ or $\lambda_3$. In this respect, to find the
closest classical states, two situations must be considered
separately.  We begin with the first case where $\lambda_{\rm max} =
\lambda_3$. The eigenvector, associated with this maximal
eigenvalue,  is $\vec{e}=(e_1=0,e_2=0,e_3=1)^t$. Reporting this
result in (\ref{minHS12}), it is simple to check that the closest
classical state (\ref{xi12}) takes the form
\begin{equation}\label{xi12-2bits+}
\chi_{12} = \frac{1}{4}\Bigg[ \sigma_{0}\otimes \sigma_{0}
+ T_{30} ~~\sigma_3 \otimes \sigma_{0} + T_{03} ~\sigma_0 \otimes \sigma_{3}
+ T_{33}~\sigma_3 \otimes \sigma_{3}\Bigg]
\end{equation}
Similarly the eigenvector associated to $\lambda_{\rm max} = \lambda_1$ is $\vec{e}=(e_1=1,e_2=0,e_3=0)^t$ and
from (\ref{minHS12}), one gets
\begin{equation}\label{xi12-2bits-}
\chi_{12} = \frac{1}{4}\Bigg[ \sigma_{0}\otimes \sigma_{0}
 + T_{03} ~\sigma_0 \otimes \sigma_{3}
+ T_{11}~\sigma_1 \otimes \sigma_{1}\Bigg].
\end{equation}
Beside the explicit derivation of closest classical states (\ref{xi12-2bits+}) and (\ref{xi12-2bits-}), another important point to be emphasized
is the relation between the matrix $K$ (\ref{matrixK12}), which encodes the geometric measure
quantum correlations in the state $\rho_{12}$, and the Bloch components of the one-qubit density matrices $\rho^{ii}$ $(i=1,2)$ and
$\rho^{ij}$ $(i\neq j)$ given respectively by (\ref{0011}) and (\ref{0110}). For this end, using the relations (\ref{recurr}),  the matrix $K$ (\ref{matrixK12}) rewrites as
\begin{equation}
K = {\rm diag}( 2 (T^{01}_1)^2 , - 2 (T^{01}_2)^2,  (T^{00}_3)^2  +  (T^{11}_3)^2 ).
\end{equation}
Furthermore, for one-qubit states $\rho^{00}$, $\rho^{01}$,
$\rho^{10}$ and $\rho^{11}$, we introduce the  analogues of the
matrix $K$ (\ref{matrixK12}). Hence, for the states $\rho^{00}$ and
$\rho^{11}$ (\ref{0011}), we introduce the $3 \times 3$ matrices
$$ K^{kk} = (0, 0, T^{kk}_3)^t ~(0, 0, T^{kk}_3), \qquad k = 0,1, $$
and similarly, we introduce the matrices
$$K^{kl} = (T^{kl}_1, iT^{kl}_2, 0)^t ~(T^{kl}_1, iT^{kl}_2, 0) \quad {\rm for} \quad (k,l) = (0,1)~{\rm or}~(1,0).$$
for the  states  $\rho^{01}$ and $\rho^{10}$ (\ref{0110}).
Subsequently, one verifies
$$ K = 2 (K^{00} + K^{01} +  K^{10} +  K^{11} ).$$
This remarkable relation holds also for the states containing three or more qubits as a consequence  of the symmetry invariance of the
multi-qubit system under consideration.  A detailed analysis of this issue is presented
in what follows.

%%%%%%%%%%%%%%%%%%%%%%%%%%%%%%%%%%%%%%%%%%%%%%%%%%%%%%%%%%%%%%%%%%%%%%%%
\subsection{Three-qubit states}
%%%%%%%%%%%%%%%%%%%%%%%%%%%%%%%%%%%%%%%%%%%%%%%%%%%%%%%%%%%%%%%%%%%%%%%%

We now face the problem of finding the pairwise quantum discord in the three qubit states of the form (\ref{rho123}). This extends the results presented
in the previous subsection. More especially, we analytically determine the  pairwise quantum discord between the qubit $1$ and
the subsystem $23$ in the state $\rho_{123}$ (\ref{rho123}) and we find the closest classical  tripartite states. To achieve this, we write the  density matrix (\ref{density})
as follows
\begin{equation}\label{3X-fano}
\rho_{123}=\frac{1}{8}\bigg[T_{000} ~\sigma_0\otimes \sigma_0\otimes \sigma_0 + T_{300}~\sigma_3\otimes
\sigma_0\otimes \sigma_0 +  \sum_{(\beta,\gamma)\neq (0,0)} T_{0 \beta \gamma}~ \sigma_{0}\otimes \sigma_{\beta}\otimes
\sigma_{\gamma} +  \sum_i \sum _{(\beta,\gamma)\neq (0,0)} T_{i \beta \gamma} ~\sigma_{i}\otimes \sigma_{\beta}\otimes
\sigma_{\gamma}\bigg]
\end{equation}
The classical states  (i.e., states presenting zero discord between the qubit $1$ and the subsystem $23$) are of the form
\begin{equation}\label{chi}
\chi_{1|23} = p_1 \vert \psi_1 \rangle \langle \psi_1 \vert \otimes \rho_1^{23} + p_2 \vert \psi_2 \rangle \langle \psi_2 \vert \otimes \rho_2^{23}
\end{equation}
where $\{ \vert \psi_1 \rangle , \vert \psi_2 \rangle \}$ is an orthonormal basis related to the qubit $1$. The density matrices  $\rho_i^{23}$ ($i=1,2$) corresponding to the subsystem $23$ write as
$$ \rho_i^{23} = \frac{1}{4} \bigg[  \sum_{\alpha,\beta} {\rm Tr}(\rho_i^{23} \sigma_{\alpha}\otimes \sigma_{\beta})\sigma_{\alpha}\otimes \sigma_{\beta}\bigg].$$
The Fano-Bloch form of the  tripartite classical state (\ref{chi}) is given by
\begin{equation}\label{chi-matrix}
\chi_{1|23} = \frac{1}{8}\Bigg[ \sigma_{0}\otimes \sigma_{0} \otimes \sigma_{0}
+ \sum_{i=1}^3 te_i ~~\sigma_i \otimes \sigma_{0} \otimes \sigma_{0} $$
$$+ \sum_{(\alpha,\beta)\neq (0,0)} (s_+)_{\alpha,\beta} ~\sigma_0 \otimes \sigma_{\alpha}\otimes \sigma_{\beta}
+ \sum_{i=1}^3  \sum_{(\alpha,\beta)\neq (0,0)}  e_i (s_-)_{\alpha,\beta} ~\sigma_i \otimes \sigma_{\alpha}\otimes \sigma_{\beta}\Bigg]
\end{equation}
where
$$ t = p_1-p_2 \qquad e_i = \langle \psi_1 \vert \sigma_i \vert \psi_1 \rangle  \qquad
(s_{\pm})_{\alpha,\beta} = {\rm Tr}\big((p_1\rho_1^{23} \pm p_2\rho_2^{23}) \sigma_{\alpha}\otimes \sigma_{\beta}\big)$$
The Hilbert-Schmidt distance between the three qubit state $\rho_{123}$ (\ref{3X-fano}) and a classical state  (\ref{chi-matrix}) is
\begin{equation}\label{HS}
|| \rho_{1|23} - \chi_{1|23}||^2 = \frac{1}{8} \bigg[  (t^2 -2te_3T_{300} + T^2_{300}) +  \sum_{(\alpha,\beta)\neq (0,0)}
(T_{0\alpha\beta} -  (s_+)_{\alpha,\beta})^2 + \sum_{i=1}^3  \sum_{(\alpha,\beta)\neq (0,0)} (T_{i\alpha\beta} -  e_i(s_-)_{\alpha,\beta})^2  \bigg]
\end{equation}
To derive the closest classical state as measured by Hilbert-Schmidt, an optimization with  respect to the parameters $t$, $e_i$ ($i=1,2,3$) and $(s_{\pm})_{\alpha,\beta}$
is performed. Thus, the minimal distance is attainable by setting zero the partial derivatives of the
Hilbert-Schmidt distance (\ref{HS})  with respect to $t$ and $(s_{\pm})_{\alpha,\beta}$.  This gives
\begin{equation}\label{solu}
t = e_3 T_{300}  \qquad (s_{+})_{\alpha,\beta} = T_{0\alpha\beta}   \qquad (s_{-})_{\alpha,\beta} = \sum_{i=1}^3 e_i T_{i\alpha\beta}
\end{equation}
Reporting the results (\ref{solu}) in (\ref{HS}), one obtains
\begin{equation}\label{HS123}
|| \rho_{1|23} - \chi_{1|23}||^2 = \frac{1}{8} \bigg[  T^2_{300} - e^2_3T^2_{300}
+ \sum_{i=1}^3  \sum_{(\alpha,\beta)\neq (0,0)} T^2_{i\alpha\beta} - \sum_{i,j=1}^3  \sum_{(\alpha,\beta)\neq (0,0)} e_ie_jT_{i\alpha\beta}T_{j\alpha\beta}\Bigg]
\end{equation}
to be optimized with respect to the three components of the unit vector $\vec{e} = (e_1, e_2, e_3)$.
The equation (\ref{HS123}) can re-expressed as
\begin{equation}\label{HS1}
|| \rho_{1|23} - \chi_{1|23}||^2 = \frac{1}{8} \big[  ||x||^2 + ||T||^2 -\vec{e}(xx^t + TT^t)\vec{e}~^t\big]
\end{equation}
in terms of the   $1\times 3$ matrix defined by
\begin{equation}\label{x123}
x^t:=(0,0,T_{300})
\end{equation}
and the $3\times 15$ matrix given by
\begin{equation}\label{T123}
T := (T_{i\alpha  \beta}) ~~~ {\rm with}~~~~ (\alpha,\beta) \neq (0,0)
\end{equation}
%and we define the   $4\times 16$ correlation matrix  writes formally
%\begin{equation}\label{matrixT}
%T=\left(%
%\begin{array}{cc}
% T_{000}  &  T_{0\alpha  \beta}= y \\
%T_{i00} = x & T= T_{i\alpha  \beta}  \\
%\end{array}%
%\right)
%\end{equation}
Setting
\begin{equation}\label{matrixK-class2}
K = xx^{t}+TT^{t},
\end{equation}
and reporting (\ref{x123}) and (\ref{T123}) in (\ref{matrixK-class2}), one obtains after some   tedious calculations
\begin{equation}\label{matrixK1}
K= {\rm diag}(k_1, k_2, k_3)
\end{equation}
where  $k_1$, $k_2$  and $k_3$ are given by
\begin{equation}\label{k1k2k3}
k_1 = \sum_{i=1,2} \sum_{j=0,3} T^2_{1ij}+ T^2_{1ji}, \quad k_2 = \sum_{i=1,2} \sum_{j=0,3} T^2_{2ij}+ T^2_{2ji},
\quad k_3 = \sum_{i=0,3} \sum_{j=0,3} T^2_{3ij} +  \sum_{i=1,2} \sum_{j=1,2}   T^2_{3ij}.
\end{equation}
%\begin{equation}\label{Kij-class2}
%K_{kl} = \sum_{i=1,2} \sum_{j=0,3} T_{kij}T_{lij}+ T_{kji}T_{lji}
%\end{equation}
%for $k,l= 1,2$ and the remaining  matrix element writes
%\begin{equation}\label{Kij-class2}
%K_{33} = \sum_{i=0,3} \sum_{j=0,3} T^2_{3ij} +  \sum_{i=1,2} \sum_{j=1,2}   T^2_{3ij}
%\end{equation}
Using the relations (\ref{relation3ret2r-class2}) and
(\ref{relation3ret2r-class22}), the eigenvalues of the matrix $K$ can be re-expressed in terms of the bipartite correlations elements
$T_{\alpha \beta}$ associated with the the qubit density matrices $\rho^{01}$, $\rho^{01}$, $\rho^{10}$
and $\rho^{11}$ (cf. (\ref{2X-fano-class2-1}) and (\ref{2X-fano-class2-2})). Therefore, one has
\begin{equation}
k_1=2[(T_{11}^{00})^2+(T_{11}^{11})^2]+4|
T^{01}_{10}|^2+4 |T^{01}_{13}|^2
\end{equation}
\begin{equation}
k_{2}= 2[(T_{22}^{00})^2+(T_{22}^{11})^2]+4|
T^{01}_{20}|^2+4 |T^{01}_{23}|^2
\end{equation}
\begin{equation}
k_{3}= 2[(T_{30}^{00})^2+(T_{30}^{11})^2] +2[(T_{33}^{00})^2+(T_{33}^{11})^2]+4|
T^{01}_{31}|^2+4 |T^{01}_{32}|^2
\end{equation}
%where we have used the relation $\overline{T^{01}_{\alpha,\beta}} = T^{10}_{\alpha,\beta}$.
%The non-diagonal element $K_{12}$ is
%\begin{equation}
%K_{12}= K_{21}=2({T}_{21}^{00}T_{11}^{00}+T_{21}^{11}{T}_{11}^{11})+2({T}_{22}^{00}T_{12}^{00}+T_{22}^{11}{T}_{12}^{11})$$
%$$+2(\overline{T^{01}_{20}}T^{01}_{10}+T^{01}_{20}\overline{T^{01}_{10}})
%+2(\overline{T^{01}_{23}}T^{01}_{13}+T^{01}_{23}\overline{T^{01}_{13}})
%\end{equation}
%For the density matrices $\sigma^{01}$ and $\sigma^{10}$, we define the analogue of Mullemer ??? matrix as
%\begin{equation}\label{matrixT}
% T^{ij}=\left(%
%\begin{array}{cc}
% 0  &   y^{ij} \\
% x^{ij} & t^{ij}  \\
%\end{array}%
%\right)
%\end{equation}
%where
%$$ y^{ij} = (T^{ij}_{01} , T^{ij}_{02} , 0)$$
%$$(x^{ij})^t = (T^{ij}_{10} , T^{ij}_{20} , 0) $$
%and%\begin{equation}\label{matrixt}
% t^{ij}=\left(%
%\begin{array}{ccc}
% 0 & 0  &  T^{ij}_{13}\\
% 0 & 0 & T^{ij}_{23}  \\
% T^{ij}_{31} & T^{ij}_{32} & 0  \\
%\end{array}%
%\right)
%\end{equation}
%and we define the following matrix
%$$K^{ij} = x^{ij} (x^{ij})^{\dagger} + t^{ij}(t^{ij})^{\dagger}$$
%For the two-qubit density matrices $\sigma^{00}$ and $\sigma^{11}$, which are $X$ shaped, the
%Mullmerer??? matrices $K^{00}$ and $K^{11}$ can be obtained from the matrix (\ref{Kij}) modulo some
%obvious substitution. It follows that the matrix $K$ can be
%written in terms of $K^{00}$, $K^{10}$, $K^{01}$ and $K^{11}$ as% \begin{equation}
%K= 2( K^{00}+K^{10}+K^{01}+K^{11})
%\end{equation}
Finally, using  (\ref{Tii}) and (\ref{Tij}), we obtain
\begin{equation}\label{L11}
k_{1} = 2 ~ \frac{(1-p^2)^2(1+p^2)}{(1+p^n\cos m\pi)^2},
\end{equation}
\begin{equation}\label{L22}
k_{2} = 2 ~ \frac{(1-p^2)^2(1+p^2)p^{2(n-3)}}{(1+p^n\cos m\pi)^2},
\end{equation}
\begin{equation}\label{L33}
k_{3} = 2 ~ \frac{(p^2+ p^{2(n-3)})(1+p^4) + 4 p^n\cos m\pi}{(1+p^n\cos m\pi)^2},
\end{equation}
%and
%\begin{equation}\label{L12}
%K_{12} =  -16 \bigg[|\sigma_{23}||\sigma_{14}|\sin(\gamma_{23} + \gamma_{14})
%+ |\sigma_{58}||\sigma_{67}|\sin(\gamma_{58} + \gamma_{67}) + |\sigma_{35}||\sigma_{17}|\sin(\gamma_{17} - \gamma_{35})
%+ |\sigma_{28}||\sigma_{46}|\sin(\gamma_{28} -\gamma_{46}) \bigg]
%\end{equation}
%The minimal Hilbert-Schmidt is obtained for $\vec{e}$ is the eigenvector of the largest eigenvalue of
%the Muller matrix given by
%$$k_1 =  \frac{1}{2} (K_{11} + K_{22}) + \frac{1}{2} \sqrt {(K_{11} + K_{22})^2 - 4(K_11K_22 - K_12K_21)}$$
%$$k_2 = \frac{1}{2} (K_{11} + K_{22}) - \frac{1}{2} \sqrt {(K_{11} + K_{22})^2 - 4(K_11K_22 - K_12K_21)}$$
%$$k_3 = K_{33}$$
The minimal value of the Hilbert-Schmidt distance (\ref{HS1}) is
reached when $\vec{e}$ is the eigenvector associated to  the largest
eigenvalue of the matrix defined by (\ref{matrixK-class2}). We
denote by $k_{\rm max}$ the largest eigenvalue among $k_1$, $k_2$
and $k_3$. Since $k_1 \geq k_2$, $k_{\rm max}$ is $k_2$ or $k_3$
depending on the number of qubits $n$ and the overlap $p$. Notice
that the sum of the eigenvalues $k_1$, $k_2$ and $k_3$ of the matrix
$K$ is exactly the sum of the Hilbert-Schmidt norm of the matrices
$x$ (\ref{x123}) and $T$ (\ref{T123})  $(i.e. k_1 + k_2 + k_3  =
||x||^2 + ||T||^2)$. It follows that the minimal Hilbert-Schmidt
distance (\ref{HS1}) writes as
\begin{equation}\label{Dg}
D_{\rm g} (\rho_{1|23}) = \frac{1}{8} (k_1 + k_2 + k_3 - k_{\rm max})
\end{equation}
and gives the geometric measure of the pairwise quantum discord in the state $\rho_{123}$ partitioned in
the subsystems $1$ and $23$. When the matrix elements of the density matrix $\rho_{123}$ (\ref{3X-fano1}) are such that
$k_{\rm max} = k_1$, one can simply verify that the closest classical state is given by
\begin{equation}\label{chi-matrix-i}
\chi^{(1)}_{1|23} = \frac{1}{8}\Bigg[ \sigma_{0}\otimes \sigma_{0} \otimes \sigma_{0}
+ \sum_{(\alpha,\beta)\neq (0,0)} T_{0\alpha\beta} ~\sigma_0 \otimes \sigma_{\alpha}\otimes \sigma_{\beta}+
\sum_{(\alpha,\beta)\neq (0,0)} T_{1\alpha\beta}   ~\sigma_1 \otimes \sigma_{\alpha}\otimes \sigma_{\beta}\Bigg]
\end{equation}
%where
%$$ T^{(i)}_{1\alpha\beta} =  \cos^2\theta_i T_{1\alpha\beta} - \cos\theta_i \sin\theta_i T_{2\alpha\beta}
%\qquad  T^{(i)}_{2\alpha\beta}  =  \sin^2\theta_i T_{2\alpha\beta} - \cos\theta_i \sin\theta_i T_{1\alpha\beta}$$
%with
%$$\tan \theta_i =\frac{K_{11} - k_{i}}{K_{12}}.$$
Conversely, in the situation where $k_{\rm max} = k_3$,  one finds
\begin{equation}\label{chi-matrix-i}
\chi^{(3)}_{1|23} = \frac{1}{8}\Bigg[ \sigma_{0}\otimes \sigma_{0} \otimes \sigma_{0}
+  T_{300} ~\sigma_3 \otimes \sigma_{0}\otimes \sigma_{0}$$
$$+ \sum_{(\alpha,\beta)\neq (0,0)} T_{0\alpha\beta}   ~\sigma_0 \otimes \sigma_{\alpha}\otimes \sigma_{\beta}
+ \sum_{(\alpha,\beta)\neq (0,0)} T_{3\alpha\beta}   ~\sigma_3 \otimes \sigma_{\alpha}\otimes \sigma_{\beta}\Bigg]
\end{equation}

%%%%%%%%%%%%%%%%%%%%%%%%%%%%%%%%%%%%%%%%%%%%%%%%%%%%%%%%%%%%%%%%%%%%%%%%
\subsection{$k$-qubit states}
%%%%%%%%%%%%%%%%%%%%%%%%%%%%%%%%%%%%%%%%%%%%%%%%%%%%%%%%%%%%%%%%%%%%%%%%

Now we come to the generalization of  the previous analysis. In this order, we shall  determine  the explicit expression of the geometric
 discord in the $k$-qubit state (\ref{rho123k-trace}) when a bipartite splitting of type  $1\vert 23\cdots k$ is considered. We also
 derive the closest classical state to the state (\ref{rho123k-trace}).  We
first expand the density matrix $\rho_{12\cdots k}$ (\ref{fanorhor123k}) as
\begin{equation}\label{3X-fanok}
\rho_{12\cdots k}=\frac{1}{2^k}\bigg[T_{00 \cdots 0} ~\sigma_0\otimes \sigma_0 \cdots \otimes \sigma_0 + T_{30 \cdots 0}~\sigma_3\otimes
\sigma_0\otimes \cdots \sigma_0 $$
$$+  \sum_{(\alpha_2, \cdots \alpha_k)\neq (0,\cdots, 0)} T_{0 \alpha_2 \cdots \alpha_k}~ \sigma_{0}\otimes \sigma_{\alpha_2}\otimes \cdots
\sigma_{\alpha_k} +  \sum_i \sum _{(\alpha_2, \cdots \alpha_k)\neq (0,\cdots, 0)} T_{i \alpha_2 \cdots \alpha_k} ~\sigma_{i}\otimes \sigma_{\alpha_2}\otimes \cdots
\sigma_{\alpha_k}\bigg]
\end{equation}
in terms of the non vanishing correlations coefficients. Any  $k$-qubit state  having zero discord is necessarily of the form
\begin{equation}\label{chik}
\chi_{1|23\cdots k} = p_1 \vert \psi_1 \rangle \langle \psi_1 \vert \otimes \rho_1^{23\cdots k} + p_2 \vert \psi_2 \rangle \langle \psi_2 \vert \otimes \rho_2^{23\cdots k}
\end{equation}
where $\{ \vert \psi_1 \rangle , \vert \psi_2 \rangle \}$ is an orthonormal basis related to the qubit $1$. The density matrices  $\rho_i^{23\cdots k}$ ($i=1,2$), corresponding to the subsystem $(23\cdots k)$,
that contains  $(k-1)$ qubits, write as
$$ \rho_i^{23\cdots k } = \frac{1}{2^{k-1}} \bigg[  \sum_{\alpha_2, \cdots \alpha_k} {\rm Tr}(\rho_i^{23\cdots k}
\sigma_{\alpha_2}\otimes \cdots \sigma_{\alpha_k})\sigma_{\alpha_2}\otimes \cdots \sigma_{\alpha_k}\bigg].$$
To examine the pairwise quantum correlations in the states (\ref{rho123k-trace}), the appropriate form for  multipartite classical state (\ref{chik}) is
\begin{equation}\label{chi-matrixk}
\chi_{1|23\cdots k} = \frac{1}{2^k}\Bigg[ \sigma_{0}\otimes \sigma_{0} \cdots \otimes \sigma_{0}
+ \sum_{i=1}^3 te_i ~~\sigma_i \otimes \sigma_{0}\cdots \otimes \sigma_{0} $$
$$+ \sum_{(\alpha_2,\cdots \alpha_k)\neq (0,\cdots, 0)} (s_+)_{\alpha_2,\cdots, \alpha_k} ~\sigma_0 \otimes \sigma_{\alpha_2}\otimes \cdots \sigma_{\alpha_k}
+ \sum_{i=1}^3  \sum_{(\alpha_2,\cdots \alpha_k)\neq (0,\cdots, 0)}  e_i (s_-)_{\alpha_2,\cdots, \alpha_k} ~\sigma_i \otimes \sigma_{\alpha_2}\otimes \cdots \sigma_{\alpha_k}\Bigg]
\end{equation}
where
$$ t = p_1-p_2 \qquad e_i = \langle \psi_1 \vert \sigma_i \vert \psi_1 \rangle  \qquad
(s_{\pm})_{\alpha_2,\cdots, \alpha_k} = {\rm
Tr}\big((p_1\rho_1^{23\cdots k} \pm p_2\rho_2^{23\cdots k})
\sigma_{\alpha_2}\otimes \cdots \sigma_{\alpha_k}\big)$$ Hence, the
Hilbert-Schmidt distance between the state $\rho_{123\cdots k}$
(\ref{3X-fanok}) and a classical state of the form
(\ref{chi-matrixk}) is given by the following expression
\begin{equation}\label{HSk}
|| \rho_{1|23\cdots k} - \chi_{1|23\cdots k}||^2 = \frac{1}{2^k} \bigg[  (t^2 -2te_3T_{30 \cdots 0} + T^2_{30\cdots 0}) +  \sum_{(\alpha_2,\cdots \alpha_k)\neq (0,\cdots, 0)}
(T_{0\alpha_2 \cdots \alpha_k} -  (s_+)_{\alpha_2,\cdots, \alpha_k})^2$$
$$ + \sum_{i=1}^3  \sum_{(\alpha_2,\cdots \alpha_k)\neq (0,\cdots, 0)} (T_{i\alpha_2 \cdots \alpha_k} -  e_i(s_-)_{\alpha_2,\cdots, \alpha_k})^2  \bigg]
\end{equation}
which must be optimized with  respect to the parameters $t$, $e_i$ ($i=1,2,3$) and
$(s_{\pm})_{\alpha_2,\cdots, \alpha_k}$ to find the closest classical states.
In this sense, we start by setting the partial derivatives of (\ref{HSk}), with respect to the parameters $t$ and $(s_{\pm})_{\alpha_2,\cdots, \alpha_k}$,  equal to zero. Thus, we get
\begin{equation}\label{soluk}
t = e_3 T_{30\cdots 0}  \qquad (s_{+})_{\alpha_2,\cdots, \alpha_k} = T_{0\alpha_2\cdots \alpha_k}   \qquad
(s_{-})_{\alpha_2,\cdots, \alpha_k} = \sum_{i=1}^3 e_i T_{i\alpha_2 \cdots \alpha_k}.
\end{equation}
Reporting the conditions (\ref{soluk}) in the expression (\ref{HSk}), one obtains
\begin{equation}\label{HSkk}
|| \rho_{1|23\cdots k} - \chi_{1|23\cdots k}||^2 = \frac{1}{2^k} \bigg[  T^2_{30\cdots 0} - e^2_3T^2_{30\cdots 0}
+ \sum_{i=1}^3  \sum_{(\alpha_2,\cdots, \alpha_k)\neq (0,0)} T^2_{i\alpha_2 \cdots \alpha_k}$$
$$ - \sum_{i,j=1}^3  \sum_{(\alpha_2,\cdots, \alpha_k)\neq (0,\cdots, 0)} e_ie_jT_{i\alpha_2 \cdots \alpha_k}T_{j\alpha_2 \cdots \alpha_k}\Bigg]
\end{equation}
that has to be optimized with respect to the three components of the unit vector $\vec{e} = (e_1, e_2, e_3)$ in order to get the minimal
Hilbert-Schmidt distance. After some algebra, the
distance  (\ref{HSkk}) takes the following compact form
\begin{equation}\label{HSkkk}
|| \rho_{1|23\cdots k} - \chi_{1|23\cdots k}||^2 = \frac{1}{2^k} \big[  ||x||^2 + ||T||^2 -\vec{e}(xx^t + TT^t)\vec{e}~^t\big]
\end{equation}
in terms of the   $1\times 3$ matrix defined by
\begin{equation}\label{x123k}
x^t=(0,0,T_{30\cdots 0})
\end{equation}
and the $3\times (4^{k-1}-1)$ matrix given by
\begin{equation}\label{T123k}
T = (T_{i\alpha_2 \cdots  \alpha_k}) ~~~ {\rm with}~~~~ (\alpha_2,\cdots, \alpha_k) \neq (0,\cdots, 0)
\end{equation}
which are the extended versions of the matrices (\ref{x123}) and (\ref{T123}) introduced for $k=3$.
Similarly to the particular cases $k=2,3$ and from the equation (\ref{HSkkk}), it is easily seen that the pairwise quantum correlation is completely characterized  by
the eigenvalues of the matrix
\begin{equation}\label{MatrixKk}
K = xx^t + TT^t.
\end{equation}
It is clear that the computation of these eigenvalues for an arbitrary multi-qubit state constitutes a very complex task. However, this complexity is considerably
reduced for  the states $\rho_{12\cdots k}$ by exploiting their
parity symmetry (i.e. commutes with $\sigma_3 \otimes \sigma_3 \cdots
\sigma_3$). This implies that the matrix $T$ (\ref{T123k})  writes formally  as
$$ T = \sum_{\alpha_2,\cdots, \alpha_k} ( T_{ 1 \alpha_2 \cdots \alpha_k} , T_{ 2 \alpha_2 \cdots \alpha_k} , 0 )^t +
\sum_{\alpha_2,\cdots, \alpha_k} ( 0, 0 , T_{ 3 \alpha_2 \cdots \alpha_k} )^t.$$
This form is more appropriate to show that the product $TT^t$ is  diagonal. The qubits forming the system described by the state  $\rho_{12\cdots k}$ are identical and invariant under
exchange symmetry. Consequently, since the  elements the density matrix  $\rho_{12\cdots k}$ are reals and in view of the recurrence relation (\ref{recur-k}), the
off-diagonal entries of the matrix $TT^t$ vanish. This result has been discussed already  for $k=2$, $k=3$ and will be explicitly
proved hereafter for $k=4$.  It follows that the matrix $K$ (\ref{MatrixKk})  is diagonal
%{\bf In computing  the elements of  the matrix $TT^t$, we use the fact that the matrix elements of $\rho_{123\cdots k}$ are reals and
%the recurrence relation (\ref{recur-k}). Thus, one can verify that
%the cross product terms of type $T_{ 1 \alpha_2,\cdots, \alpha_k} T_{ 2 \alpha_2,\cdots, \alpha_k}$ vanish. In other words,
%if $T_{ 1 \alpha_2,\cdots, \alpha_k}$ is non zero, the element $T_{ 2 \alpha_2,\cdots, \alpha_k}$ vanishes and vice-versa. This is imposed
%by the invariance under parity symmetry.}
%It follows that the matrix $TT^t$ is diagonal and we have
$$ K = {\rm diag} (k_1 , k_2 , k_3)$$
where
$$ k_1 = \sum_{\alpha_2,\cdots, \alpha_k} T^2_{ 1 \alpha_2 \cdots \alpha_k}$$
$$  k_2 = \sum_{\alpha_2,\cdots, \alpha_k} T^2_{ 2 \alpha_2 \cdots \alpha_k} $$
$$ k_3 = T^2_{30\cdots 0} + \sum_{\alpha_2,\cdots, \alpha_k\neq 0} T^2_{ 3 \alpha_2 \cdots  \alpha_k}$$
To exemplify this procedure, we consider the situation where $k=4$. In this case, the $3\times 63$ matrix elements
of $T$ defined by (\ref{T123k}) can be explicitly derived  using the equation
(\ref{recu-4}). A straightforward but lengthy computation  shows that the $3 \times 3$ matrix $K$ is diagonal
and the corresponding  eigenvalues are
\begin{equation}\label{k1-4}
k_1 =  \sum_{k = 0,3 }\sum_{i=1,2} \sum_{j=0,3} T^2_{1kji} + \sum_{k =  1,2}\sum_{j=1,2} \sum_{i=1,2} T^2_{1kji}$$
$$ +\sum_{k = 0,3 }\sum_{j=1,2} \sum_{i=0,3} T^2_{1kji} + \sum_{k = 1,2}\sum_{j=0,3} \sum_{i=0,3} T^2_{1kji}
\end{equation}
\begin{equation}\label{k2-4}
k_2 =  \sum_{k = 0,3 }\sum_{i=1,2} \sum_{j=0,3} T^2_{2kji} + \sum_{k =  1,2}\sum_{j=1,2} \sum_{i=1,2} T^2_{2kji}$$
$$ + \sum_{k = 0,3 }\sum_{j=1,2} \sum_{i=0,3} T^2_{2kji} + \sum_{k = 1,2}\sum_{j=0,3} \sum_{i=0,3} T^2_{2kji}
\end{equation}
\begin{equation}\label{k3-4}
k_3 =   \sum_{k = 0,3 }\sum_{i=0,3} \sum_{j=0,3} T^2_{3kji} + \sum_{k = 1,2}\sum_{j=1,2} \sum_{i=0,3} T^2_{3kji}$$
$$ + \sum_{k = 0,3 }\sum_{j=1,2} \sum_{i=1,2} T^2_{3kji} + \sum_{k =1,2}\sum_{j=0,3} \sum_{i=1,2} T^2_{3kji}
\end{equation}
%To establish the additivity relation between the matrix $K$ and the matrices $K^{kl}$ associated
%with the three-qubit density matrices $\rho^{kl}_{123}$ (\ref{rhokl123}), we use the recurrence relations (\ref{recu-4}). This
%gives
%\begin{equation}\label{k1-4}
%k_1 = 2\bigg[ \sum_{k = 0,3 }\sum_{j=1,2} ( (T^{00}_{1kj})^2 + (T^{11}_{1jk})^2 ) + \sum_{k =  1,2} \sum_{j=0,3} ( (T^{00}_{1kj})^2 + (T^{11}_{1jk})^2 )\bigg]$$
%$$ +4 \bigg[\sum_{k = 0,3 } \sum_{j=0,3} T^{01}_{1kj}T^{10}_{1kj} + \sum_{k = 1,2}\sum_{j=1,2} T^{01}_{1kj}T^{10}_{1kj}\bigg]
%\end{equation}
%\begin{equation}\label{k2-4}
%k_2 = 2\bigg[ \sum_{k = 0,3 }\sum_{j=1,2} ( (T^{00}_{2kj})^2 + (T^{11}_{2jk})^2 ) + \sum_{k =  1,2} \sum_{j=0,3} ( (T^{00}_{2kj})^2 + (T^{11}_{2jk})^2 )\bigg]$$
%$$ +4 \bigg[\sum_{k = 0,3 } \sum_{j=0,3} T^{01}_{2kj}T^{10}_{2kj} + \sum_{k = 1,2}\sum_{j=1,2} T^{01}_{2kj}T^{20}_{1kj}\bigg]
%\end{equation}
%\begin{equation}\label{k3-4}
%k_3 = 2\bigg[ \sum_{k = 0,3 }\sum_{j=0,3} ( (T^{00}_{3kj})^2 + (T^{11}_{3jk})^2 ) + \sum_{k = 1,2} \sum_{j=1,2} ( (T^{00}_{3kj})^2 + (T^{11}_{3jk})^2 )\bigg]$$
%$$ +4 \bigg[\sum_{k = 0,3 } \sum_{j=1,2} T^{01}_{3kj}T^{10}_{3kj} + \sum_{k = 1,2}\sum_{j=0,3} T^{01}_{3kj}T^{10}_{3kj}\bigg]
%\end{equation}
%Using the expressions of the eigenvalues of the Muller matrix for a
%mixed three-qubit state, one shows that
%$$ K = 2(K^{00} + K^{01} + K^{10} + K^{11})$$
The expressions (\ref{k1-4}), (\ref{k2-4}) and (\ref{k3-4})  can be simplified further. Indeed,  from the relations
(\ref{recur-k}),  which reproduce the expressions (\ref{recu-4}) for $k=4$, one obtains
\begin{equation}\label{k11-4}
k_1 = 2\bigg[ \sum_{k = 0,3 }\sum_{j=1,2} ( (T^{00}_{1kj})^2 + (T^{11}_{1jk})^2 ) + \sum_{k =  1,2} \sum_{j=0,3} ( (T^{00}_{1kj})^2 + (T^{11}_{1jk})^2 )\bigg]$$
$$ +4 \bigg[\sum_{k = 0,3 } \sum_{j=0,3} T^{01}_{1kj}T^{10}_{1kj} + \sum_{k = 1,2}\sum_{j=1,2} T^{01}_{1kj}T^{10}_{1kj}\bigg]
\end{equation}
\begin{equation}\label{k21-4}
k_2 = 2\bigg[ \sum_{k = 0,3 }\sum_{j=1,2} ( (T^{00}_{2kj})^2 + (T^{11}_{2jk})^2 ) + \sum_{k =  1,2} \sum_{j=0,3} ( (T^{00}_{2kj})^2 + (T^{11}_{2jk})^2 )\bigg]$$
$$ +4 \bigg[\sum_{k = 0,3 } \sum_{j=0,3} T^{01}_{2kj}T^{10}_{2kj} + \sum_{k = 1,2}\sum_{j=1,2} T^{01}_{2kj}T^{20}_{1kj}\bigg]
\end{equation}
\begin{equation}\label{k31-4}
k_3 = 2\bigg[ \sum_{k = 0,3 }\sum_{j=0,3} ( (T^{00}_{3kj})^2 + (T^{11}_{3jk})^2 ) + \sum_{k = 1,2} \sum_{j=1,2} ( (T^{00}_{3kj})^2 + (T^{11}_{3jk})^2 )\bigg]$$
$$ +4 \bigg[\sum_{k = 0,3 } \sum_{j=1,2} T^{01}_{3kj}T^{10}_{3kj} + \sum_{k = 1,2}\sum_{j=0,3} T^{01}_{3kj}T^{10}_{3kj}\bigg].
\end{equation}
in terms of  the three qubit correlation elements $T^{kl}_{\alpha\beta\gamma}$ associated with the  density matrices $\rho^{kl}_{123}$ (\ref{rhokl123}).
The tripartite correlations coefficients $T^{kl}_{\alpha\beta\gamma}$  are evaluated
using the recurrence relations of type (\ref{relation3ret2r-class2}) and (\ref{relation3ret2r-class22}) (modulo some obvious substitution) as expansion of bipartite correlations
associated with two qubit subsystems. Subsequently, one finds
\begin{equation}
k_1 = 16 {\cal N}^4 (1 - p^2)(1 - p^{6})\label{lambda14}
\end{equation}
\begin{equation}
k_2 = 16 {\cal N}^4 (1 - p^2)(1 -
p^{6})p^{2(n-4)}\label{lambda24}
\end{equation}
\begin{equation}
k_3 = 16 {\cal N}^4 \bigg[(1 + p^{6})(p^2 + p^{2(n-4)}) + 4
 p^n \cos m\pi\bigg]\label{lambda34}
\end{equation}
Clearly, the derivation of pairwise quantum discord in $k$-qubit mixed states between one qubit and
the other $(k-1)$ qubits, viewed as a single subsystem,  requires tedious analytical manipulation. However, it must be noticed that the party
containing $(k-1)$-qubits can be mapped onto  two logical qubits.  This encoding scheme was recently
considered in \cite{daoud2,DaoudPLA,DaoudIJQI} to examine the pairwise quantum correlations in multi-qubit systems. In this spirit,
we shall compare in the following section the geometric measure of quantum discord obtained in each picture.

%%%%%%%%%%%%%%%%%%%%%%%%%%%%%%%%%%%%%%%%%%%%%%%%%%%%%%%%%%%%%
\section{Pairwise encoding}
%%%%%%%%%%%%%%%%%%%%%%%%%%%%%%%%%%%%%%%%%%%%%%%%%%%%%%%%%%%%

Different suitable splitting scenarios are possible in investigating quantum correlations in a $n$-qubit system.
In the previous sections, we essentially focused on the quantum correlation in  $k$-qubits states $(k = 0, 1, \cdots, n-1)$ extracted by a trace
procedure from the whole system, by splitting  the system of $k$ qubits
in a single qubit and a cluster of $(k-1)$ qubits.  In this section, we shall consider the scenario where the
information contained in  the  cluster of $(k-1)$ particles is encoded in two logical qubits
$\{\vert  0\rangle_{23\cdots k} ,\vert 1\rangle_{23\cdots k} \}$ defined by
\begin{equation}
\vert\eta, \eta, \cdots \eta \rangle \equiv  b_+ \vert  0 \rangle_{23\cdots k} + b_- \vert 1 \rangle_{23\cdots k}
\qquad \vert -\eta, -\eta, \cdots, -\eta \rangle \equiv b_+ \vert 0 \rangle_{23\cdots k} - b_-
 \vert  1 \rangle_{23\cdots k} ~,\label{base23k}
\end{equation}
where $$b_{\pm} = \sqrt{\frac{1 \pm p^{k-1}}{2}} .$$
In this encoding scheme, the density matrix $\rho_{1\vert 23\cdots k} \equiv \rho_{1(23\cdots k)}$ (\ref{rho123k-trace}) rewrites, in the basis
$\{ \vert 0 \rangle \otimes \vert  0\rangle_{23\cdots k}, \vert 0 \rangle \otimes \vert  1\rangle_{23\cdots k}
,\vert 1 \rangle \otimes \vert  0\rangle_{23\cdots k},\vert 1 \rangle \otimes \vert  1\rangle_{23\cdots k} \}$, as
\begin{equation}
\rho_{1(23\cdots k)} = 2{\cal N}^2 \left( \begin{smallmatrix}
a_+^2b_+^2(1+q_k \cos m\pi)& 0
    & 0& a_+a_-b_+b_-(1+q_k\cos m\pi
)\\
0  & a_+^2b_-^2(1-q_k\cos m\pi ) & a_+a_-b_+b_-(1-q_k\cos
m\pi
) & 0 \\
0  & a_+a_-b_+b_-(1-q_k\cos m\pi ) & a_-^2b_+^2(1-q_k \cos
m\pi
) & 0 \\
a_+a_-b_+b_-(1+q_k \cos m\pi ) & 0 & 0 & a_-^2b_-^2(1+q_k \cos
m\pi)
\end{smallmatrix}
\right), \label{rho23k}
\end{equation}
or equivalently, in the Fano-Bloch representation,  as
\begin{equation}
\rho_{1(23\cdots k)} = \frac{1}{4}\sum_{\alpha \beta} R_{\alpha \beta}
\sigma_{\alpha}\otimes \sigma_{\beta}
\end{equation}
where the non vanishing matrix elements $R_{\alpha \beta}$ $(\alpha,
\beta = 0,1,2,3)$ are given by
$$ R_{00} = 1, \quad R_{11} = 2{\cal N}^2 \sqrt{(1- p^2)(1- p^{2(k-1)})}, \quad R_{22} = -2{\cal N}^2 \sqrt{(1- p^2)(1- p^{2(k-1)})}~p^{n-k}\cos
m\pi,$$ $$ R_{33} = 2{\cal N}^2 (p^k + p^{n-k}\cos m\pi), \quad
R_{03} = 2{\cal N}^2 (p^{k-1} + p^{n-k+1}\cos m\pi), \quad R_{30} =
2{\cal N}^2 (p + p^{n-1}\cos m\pi).$$
Following the standard procedure to derive the geometric discord
for a two qubit system, it is simple to check that
\begin{equation}
D_{\rm g}(\rho_{1(23\cdots k)}) = \frac{1}{4}~ {\rm min}\{ l_1 + l_2 ,
l_1 + l_3 , l_2 + l_3\}.
\end{equation}
where
$$l_1 = R_{11}^2, \quad  l_2 = R_{22}^2, \quad l_3 = R_{30}^2 + R_{33}^2.$$
Explicitly, the quantities $\lambda_1$, $\lambda_2$ and $\lambda_3$
 are given by
\begin{equation}
l_1 = 4 {\cal N}^4 (1 - p^2)(1 - p^{2(k-1)})\label{lambda1}
\end{equation}
\begin{equation}
l_2 = 4 {\cal N}^4 (1 - p^2)(1 -
p^{2(k-1)})p^{2(n-k)}\label{lambda2}
\end{equation}
\begin{equation}
l_3 = 4 {\cal N}^4 \bigg[(1 + p^{2(k-1)})(p^2 + p^{2(n-k)}) + 4
 p^n \cos m\pi\bigg]\label{lambda3}.
\end{equation}
It is remarkable that for $k=2$, $k=3$ and $k=4$, one recovers the results
(\ref{GD12lambda1},\ref{GD12lambda2},\ref{GD12lambda3}), (\ref{L11},\ref{L22},\ref{L33}) and (\ref{lambda14},\ref{lambda24},\ref{lambda34}) respectively (up to the overall multiplicative factor $2^{k-2}$). Indeed, we have
$$ D_{\rm g}(\rho_{1(23\cdots k)}) =  \frac{1}{2^{k-2}} D_{\rm g}(\rho_{1\vert 23\cdots k}). $$
This
shows that encoding $(k-1)$-qubit in two logical qubits constitutes an alternative and efficient way to compute easily the
geometric measure of quantum discord.

%Finally, the global geometric discord in the multi-qubits Schro\"odinger cat state (\ref{ncs}) is given by

%\begin{equation}
%D_{\rm g} = \sum_{k=2}^n
%\end{equation}

%%%%%%%%%%%%%%%%%%%%%%%%%%%%%%%%%%%%%%%%%%%%%%%%%%%%%%%%%%%%%%%%%%%%%%%%%%%%%%%%%%%%%%%%%%%
\section{ Concluding remarks}
%%%%%%%%%%%%%%%%%%%%%%%%%%%%%%%%%%%%%%%%%%%%%%%%%%%%%%%%%%%%%%%%%%%%%%%%%%%%%%%%%%%%%%%%%%%

In this paper, we developed a general algorithm to evaluate
the pairwise geometric discord
in a mixed state $\rho_{123\cdots k}$ comprised $k$-qubits.
This provides a closed analytical expressions for the geometric quantum discord based
on Hilbert-Schmidt distance. We especially considered multi-qubit states possessing parity invariance and exchange symmetry.
A detailed analysis is performed for reduced density matrices $\rho_{123\cdots k}$
obtained by a trace procedure from a balanced  superpositions of
 symmetric $n$-qubit states. Two  splitting schemes were discussed. In the first one, where the reduced density is denoted by
 $\rho_{1\vert 23\cdots k}$, a recursive algorithm is proposed  to determine explicitly the pairwise geometric discord between the first qubit and
 the remaining $(k-1)$ qubits.  The  parity  and exchange symmetries simplify considerably the
 determination of the geometric measure of quantum discord. The recursive approach offers a very usefull prescription to determine geometric quantum discord
 in terms of two qubits correlations matrices.  This constitutes the key ingredient in deriving the geometric discord. Another important issue we examined in this
 work concerns the explicit derivation  of classical (zero discord) states. We have also shown that there exists an alternative scheme offering a
simple procedure to get the compute the geometric discord. This uses a bipartition scheme according to which the system grouping the $(k-1)$-qubits in the state
$ \rho_{123\cdots k}$ is mapped into a set of two logical qubits. Remarkably the two schemes lead to the same result for the Hilbert-Schmidt measure of  pairwise geometric discord.

We believe that the results obtained in this work can be extended to other classes of
multi-qubit states. We also notice  that they can be exploited in evaluating  multipartite geometric quantum discord
 in the spirit of the results recently obtained in
\cite{Ma}. Finally, another interesting application of the results obtained here, that deserve a special attention, concerns the distribution of geometric quantum discord
between the different components of a multi-qubit system.

\end{document}